\documentclass[12pt]{JHEP3}
\usepackage{epsfig}
\preprint{ {\tt hep-th/0606078} }
\newcommand{\be}[1]{ \begin{equation}\label{#1} }
\newcommand{\ee}{\end{equation}}
\newcommand{\bea}[1]{\begin{eqnarray}\label{#1} }
\newcommand{\eea}{\end{eqnarray}}
\newcommand{\eq}[1]{(\ref{#1})}

\newcommand{\ra}{\rangle}
\newcommand{\la}{\langle} 
\newcommand{\sn}{{\rm sn}}
\newcommand{\cn}{{\rm cn}}
\newcommand{\dn}{{\rm dn}}
\newcommand{\p}{{\wp}}
\newcommand{\e}{\eta}
\newcommand{\eb}{\bar{\eta}}
\newcommand{\me}{|\eta|}
\newcommand{\mone}{|1-\eta |}
\newcommand{\s}{\sigma}

\title{From Spacetime to Worldsheet: Four point correlators}
\author{Justin R. David, Rajesh Gopakumar \\
Harish-Chandra Research Institute, \\
Chhatnag Road., Jhunsi, \\
Allahabad 211019, India.
\\
\email{justin, gopakumr@hri.res.in}
}

\abstract{The Schwinger representation gives a systematic 
procedure for recasting large $N$ field theory amplitudes
as integrals over closed string moduli space.  
This procedure has recently been applied 
to a class of free field  four point functions 
by Aharony, Komargodski and Razamat, to study the leading terms in the 
putative worldsheet OPE.  
Here we observe that the dictionary between
Schwinger parameters and the cross ratio of the four punctured sphere
actually yields an explicit expression for the full 
worldsheet four point correlator in many such cases. This expression 
has a suggestive form and obeys various properties,
such as 
crossing symmetry and mutual locality,
expected of a correlator in a two dimensional CFT. 
Therefore one may
take this to be a candidate four point function
in a worldsheet description of closed strings on highly curved 
$AdS_5\times S^5$. The general framework, that we develop for 
computing the relevant Strebel differentials, also admits a 
systematic perturbation expansion which 
would be useful for studying more general
four point correlators.}

\begin{document}
\baselineskip 4ex

\section{Introduction}

One of the major hurdles, which repeatedly crops up, in the study of  
the AdS/CFT conjecture \cite{Maldacena:1997re, Gubser:1998bc, Witten:1998qj}
is the lack of a complete worldsheet description 
of the closed string theory on $AdS_5\times S^5$. The conventional RNS
formulation is not of much use because of the presence of RR flux in the 
background. The Green-Schwarz approach, on the other hand,
has the problem of requiring
some kind of lightcone gauge fixing to make it tractable \cite{Metsaev:1998it,Metsaev:2000yf,Metsaev:2000yu}, which in turn 
leads to subtleties in computing scattering amplitudes.There also exists a covariant formulation of the worldsheet theory
due to Berkovits \cite{Berkovits:2000fe, Berkovits:2000yr}, 
but it has not been developed to the stage where we can use 
it for comparisons with gauge theory calculations, specially 
in the perturbative domain of weak 'tHooft coupling\footnote{Many different 
approaches have been taken to study the string dual of the weakly coupled gauge theory.
For a light cone approach, see \cite{Bardakci:2001cn, Thorn:2002fj, 
Bardakci:2002xi, Gudmundsson:2002jf, Thorn:2003ux, Bardakci:2005kv}.
For a string bit approach, see \cite{Haggi-Mani:2000ru, Verlinde:2002ig,
Zhou:2002mi, Vaman:2002ka} and also from a discretised worldsheet
point of view \cite{Dhar:2003fi, Okuyama:2004bd, Alday:2005nd, 
Engquist:2005yt, Alday:2005kq, Karch:2002vn, Clark:2003wk}.
See also \cite{Bonelli:2004ve, Bianchi:2003wx, Beisert:2003te}.}.

Perhaps, what these difficulties are pointing to, is that quantising the
worldsheet theory of strings in these backgrounds may be a bit like   
quantising Liouville theory (For a review, see \cite{Teschner:2001rv}). 
The path integral (as well as the operator)
approach have yielded limited information about Liouville theory. The most 
complete description, yet, has come from the algebraic approach -- the 
so-called conformal bootstrap. This came in the form of inspired 
conjectures for the three point functions \cite{Dorn:1994xn,  Zamolodchikov:1995aa}.
It was then understood that
the crossing symmetry constraints on the structure constants, coming from four point functions, 
was sufficent for a purely algebraic determination \cite{Teschner:1995yf, 
Pakman:2006hm}.

In like manner, we could aim for a purely algebraic characterisation
of the full worldsheet CFT (with ghosts and/or other auxiliary fields)
for $AdS_5\times S^5$.
The four point functions of this $c=0$ CFT would then contain the 
information necessary to characterise the theory. By factorising in 
different channels one can read off the structure constants  $C_{ijk}$ and the structure
of the conformal blocks. 

How can we obtain these four point functions 
of the worldsheet CFT, without much knowledge about 
the sigma model? A concrete prescription to implement open-closed string 
duality for large $N$ gauge theories has been given in
 \cite{Gopakumar:2003ns,Gopakumar:2004qb, Gopakumar:2004ys, Gopakumar:2005fx}.
Essentially, one recasts field theory correlation 
functions, in Schwinger parametrisation,
as integrals over the closed string moduli space of 
worldsheet correlators (see  also \cite{Carfora:2004fd, Akhmedov:2004yb, Mamedov:2005uw, Furuuchi:2005qm, Gili:2006pb}  for 
some further explorations). In particular,
one can study the planar four point function of gauge invariant operators
in the free gauge theory (as a starting point for a perturbative 
expansion in the 't Hooft coupling $\lambda$). The prescription 
of \cite{Gopakumar:2004qb,Gopakumar:2005fx}
enables one to rewrite the simple spacetime answer
as an integral over the cross ratio 
that parametrises the four punctured sphere. The integrand is 
then a natural
candidate\footnote{The answer given by this prescription is presumably that 
obtained by fixing the worldsheet diffeomorphisms to the locally flat metric given by the 
Strebel construction of the Riemann surface (see \cite{Gopakumar:2005fx}). Of course, one may always add total derivatives 
on moduli space  to the answer obtained this way.}  
for the four point function of the corresponding vertex 
operators in the worldsheet CFT. It is, in fact, a 
{\it nontrivial} check   
of the proposal that the integrand thus obtained
(expressed in terms of the cross ratio) even satisfies all the 
generic properties
one might expect of a worldsheet correlator in a local 2d CFT. 
This is certainly not guaranteed by the construction. 

There is, however, a technical hurdle to be overcome in
implementing and  checking 
this proposal. Generically, the change of variables between the 
Schwinger parameters and the usual complex parameters on moduli space 
(such as the cross ratio, in the case of the four punctured sphere)
is given by a transcendental relation \cite{Gopakumar:2005fx} which makes it hard to obtain
explicit expressions for the worldsheet correlator. Therefore one
strategy is to look at special correlators for which there is a 
simplification. 

In \cite{Aharony:2006th} four point correlation functions of the form
\be{Yfourpt}
\Gamma^{(4)}_{ \{J_i\} }(x_1,x_2,x_3)=\la {\rm Tr}\Phi^{J_1}(x_1){\rm Tr}\Phi^{J_2}(x_2)
{\rm Tr}\Phi^{J_3}(x_3){\rm Tr}\Phi^{J}(0) \ra
\ee
were considered (with $J=J_1+J_2+J_3$ and $\Phi$ being an adjoint field 
under the $U(N)$ gauge group). In these cases, because of the simpler
nature of the free field contractions  it is possible 
to be much more explicit. In fact, Aharony et. al \cite{Aharony:2006th} 
considered the contribution to \eq{Yfourpt}  from the $Y$ shaped diagram 
of free field theory and used
the change of variables to 
study the behavior of the candidate worldsheet correlator when one of the 
punctures approaches another. In other words, they looked at the 
leading (and next to leading) behaviour in the worldsheet OPE. They 
found that for both these terms the behaviour was consistent with what might  
be expected of a worldsheet correlator in a string theory. Though 
the worldsheet conformal weights $h,\bar{h}$ could be 
individually half integral, the difference $(h-\bar{h})$ was an integer, at least to this order.

In this paper, we will develop a general method for obtaining the change 
of variables from the Schwinger parameters of a generic free field 
four point function to the cross ratio $\e$. 
Since the dictionary between the Schwinger parameters 
and $\e$ is through Strebel quadratic 
differentials on the four punctured sphere, we need to study the 
latter. Generically the expressions will involve
elliptic functions. However, 
as expected, things simplify for the special class of  
$Y$ shaped Feynman graphs studied by Aharony et.al. \cite{Aharony:2006th}. On
taking a particular limit, our method recovers their results. 
In fact, by working directly 
in terms of the cross ratio $\e$ we are able to find    
a completely explicit expression for the corresponding worldsheet correlator. 

To relate to results for ${\cal N}=4$ Yang-Mills theory, 
we could study the analogue of \eq{Yfourpt} which is 
\be{Extr}
\tilde{\Gamma}^{(4)}_{ \{J_i\} }(x_1,x_2,x_3) =\la {\rm Tr} Z^{J_1}(x_1){\rm Tr} Z^{J_2}(x_2)
{\rm Tr} Z^{J_3}(x_3){\rm Tr}\bar{Z}^{J}(0)  \ra .
\ee
Here and below $X, Y,Z$ will denote 
the complex scalars in the three chiral multiplets of ${\cal N}=4$ Yang-Mills theory.
With $J=J_1+J_2+J_3$, \eq{Extr} is an example of a so-called extremal 
correlator \cite{D'Hoker:1999ea}\footnote{There is a large amount of 
AdS/CFT literature on four point functions of various operators. See the reviews
\cite{Aharony:1999ti}\cite{D'Hoker:2002aw} for references.}. 
In fact, these are a particularly interesting class of correlators to study from
the point of view of the AdS/CFT conjecture since 
they are not renormalised from their free field value. This is a generalisation of the 
non-renormalisation theorem for three point functions \cite{Lee:1998bx}. 
The $Y$ shaped diagram is one of the contributions to this correlator in the free theory
and thus the expression we derive is of relevance for this case. 
However, there are other diagrams that need to be taken into account as well,
in evaluating 
the complete worldsheet correlator\footnote{\label{aharony} We are grateful to O. Aharony,
Z. Komargodski and S. Razamat
for pointing this out to us. The contribution of these additional diagrams are 
currently being evaluated\cite{Ofer}.}. 

However, we can readily give 
examples of   spacetime correlators which gets contributions in the free
${\cal N}=4$ Yang-Mills theory {\it only} from the $Y$ shaped graph.
\be{Yeta}  
\Gamma^{(4)}_{ \{J_i\} }(x_1,x_2,x_3) =\la {\rm Tr} X^{J_1}(x_1){\rm Tr} Y^{J_2}(x_2)
{\rm Tr }Z^{J_3}(x_3){\rm Tr}(\bar{X}^{J_1}\bar{Y}^{J_2}\bar{Z}^{J_3})(0)  \ra 
\ee
The operators entering here are not all chiral primaries unlike in \eq{Extr}. The last operator is an admixture of stringy and supergravity modes in the dual theory. 
We can use the change of variables mentioned above to express this correlator as a worldsheet  
integral
\be{gameta}
\Gamma^{(4)}_{ \{J_i\} }(x_1,x_2,x_3)= \int d^2\e G^{\{J_i\}}_{\{x_i\}}(\e, \eb).
\ee
One of the results of this paper 
is that  the above change of variables applied to this amplitude  yields an 
explicit expression for the candidate worldsheet correlator. 
$G^{\{J_i\}}_{\{x_i\}}(\e, \eb)$ in \eq{Yeta}
\newpage
\bea{Yresult}
&&G^{\{J_i\}}_{\{x_i\}}(\e, \eb)
=C(J_i){(1+\me +\mone)^{1\over 2}\over \me \mone}\times  \qquad \qquad \qquad \qquad \qquad 
\qquad \qquad \qquad \qquad \cr
&&{(1-\me+\mone)^{J_1-{1\over 2}}(1+\me-\mone)^{J_2-{1\over 2}}
(-1+\me +\mone)^{J_3-{1\over 2}}\over [x_1^2(1-\me+\mone)+x_2^2(1+\me-\mone)
+x_3^2(-1+\me +\mone)]^J}.
\eea

If we are to make the identification
\be{wscorr}
G^{\{J_i\}}_{\{x_i\}}(\e, \eb)=\la {\cal V}^{J_1}_{x_1}(0){\cal V}^{J_2}_{x_2}(1)
{\cal V}^{J_3}_{x_3}(\infty)
\bar{{\cal V}}^{\{J\}}_{x_4=0}(\e, \eb) \ra_{WS},
\ee
with a worldsheet correlator of primary vertex operators, 
then an essential  requirement is that $G^{\{J_i\}}_{\{x_i\}}(\e, \eb)$ satisfy the 
crossing symmetry 
relations
\bea{crossing}
G^{J_2 J_1 J_3 J}_{x_2 x_1 x_3}(1-\e,1-\eb)&=&
G^{J_1 J_2 J_3 J}_{x_1 x_2 x_3}(\e,\eb),\cr
G^{J_3 J_2 J_1 J}_{x_3 x_2 x_1}({1\over \e},{1\over \eb})&=&
\me^4 G^{J_1 J_2 J_3 J}_{x_1 x_2 x_3}(\e,\eb).
\eea
As can be readily verified, $G^{\{J_i\}}_{\{x_i\}}(\e, \eb)$ in \eq{Yresult}
satisfies these relations. It is also consistent with locality 
as all the terms in the OPE (when $\e\rightarrow 0$) 
\be{ope}
G^{\{J_i\}}_{\{x_i\}}(\e, \eb)=\sum_{h,\bar{h}}C^{J_i,x_i}_{h,\bar{h}}\e^h\eb^{\bar{h}}
\ee
have $(h-\bar{h})$ 
integral, even though the weights can be individually half integral. 
This generalises the results of \cite{Aharony:2006th}, for the 
leading couple of terms,
to all orders in the expansion. 

The form of the correlator is also very suggestive with the dependence on 
$\me$ and $\mone$ being what one might expect of a correlation function 
of local operators inserted at $0,1,\infty$ and $\e$. As we will discuss 
in Sec.4, there is a relation of individual terms in this expression
to the four point correlators of spin fields. This raises the possibility 
of understanding at least this class of correlators in 
an elementary way. It should be mentioned that 
the procedure by which \eq{Yresult} is obtained, readily generalises to any 
other correlator which gets a contribution from a $Y$ shaped diagram.
One can easily write 
the corresponding 
worldsheet expression for any such gauge correlator in the ${\cal N}=4$ Yang-Mills theory,
For that matter, the theory need not even be supersymmetric.

We will also go ahead and consider a systematic perturbation expansion
around the limiting form of the Strebel differential we had employed
in the above considerations. 
This expansion would be relevant to studying correlators, more general
than the one considered above. The expansion is surprisingly non-trivial  
and needs to be done quite carefully. We will give partially
explicit expressions
for the perturbed cross ratio in terms of the Strebel lengths (which
are identified with the Schwinger parameters).    
We will however postpone an application of this procedure 
to studying more general worldsheet 
correlators for the future.

The organisation of this paper is as follows. In the next section we 
review some general facts about Strebel differentials and in particular, 
the parametrisation of the differential for the four punctured sphere.
We will then give the general procedure for obtaining the relation between the 
Schwinger parameters and the cross ratio. In section 3, we show how 
this change of variables simplifies in a particular limit which corresponds 
to the $Y$ diagrams that contribute to the correlators defined above. 
We obtain the explicit relation between the cross ratio and the 
Schwinger parameters (see \eq{etaso}). 
Those who would like to skip the various 
technicalities of Strebel differentials can directly go to \eq{etaso}. 
We use 
this relation in section 4. to study the 
Schwinger parametrised form of the 
correlator \eq{Yeta}. We deduce the form \eq{Yresult}
and comment on its various features as well as the clues it might give 
us about the worldsheet theory. In section 5., we give an algorithm for performing a  
systematic perturbation around the limiting differential of section 3. We 
obtain a well defined expansion in terms  of the Strebel lengths in certain limits.  Section 6. is a brief conclusion. Various appendices contain technical details that arise
 at different points in the main text.

\section{Strebel differentials and the four punctured sphere}

\subsection{Strebel differentials and Schwinger parameters}
 
The precise recipe for obtaining the closed string correlators from the 
Schwinger representation of the field theory amplitude is via a special
kind of holomorphic quadratic differential $\phi(z)dz^2$
on the corresponding 
closed string Riemann surface. The properties of these Strebel differentials
have been reviewed in \cite{Gopakumar:2005fx, Aharony:2006th} and we refer 
the reader to 
these papers and references therein. 

Very briefly,  
Strebel differentials have double 
poles at $n$ marked points (which will be 
identified with closed string vertex operator
insertions). 
There is a critical graph (of genus $g$ with $n$ faces) 
associated to each such differential with each face of the critical 
graph enclosing a double pole of the Strebel differential. 
The vertices of this graph are 
the zeroes of the differential, with a vertex of valence $k$ associated to
a zero of order $(k-2)$. The edges of the graph are the so-called non-closed 
horizontal trajectories along which $\sqrt{\phi(z)}dz$ is real. 
Therefore the Strebel lengths defined by 
\be{streblngth}
l_r=\int_{e_r}\sqrt{\phi(z)}dz
\ee
are real, where $e_r$ is an edge connecting 
two zeroes of the differential.

As can be guessed from the identification of the poles (in each face)
with 
vertex operator insertions, the {\it dual} 
to the critical graph  is associated with the field theory 
Feynman diagrams in the $AdS/CFT$ correspondence. 
More precisely,  the skeleton diagram \cite{Gopakumar:2004qb},
obtained 
by gluing together homotopic Wick contractions in Feynman diagrams, 
is dual to the critical graph \cite{Gopakumar:2005fx}. Thus, for
a Feynman diagram of genus $g$ with $n$ 
vertices, we need to consider a Strebel differential on a Riemann 
surface of the same genus and with $n$ marked points. It is a theorem 
due to Strebel that given such a Riemann surface $\Sigma_{g,n}\in 
{\cal M}_{g,n}$ {\it together with} $n$ residues $\{p_a\}$ at the double poles,
there is a unique Strebel differential. As a consequence,
there is a one to one mapping between the $(3g-3+n)$
complex moduli $\e_i$ and $n$ residues
$p_a$ to the $(6g-6+3n)$ Strebel lengths $l_r$.
Summing over the inequivalent skeleton graphs of genus $g$ with $n$ vertices,
and varying the lengths of the edges then gives a single cover of the decorated moduli space
${\cal M}_{g,n}\times R_+^n$

The proposal to implement open-closed duality is to identify 
the inverse Schwinger times
\cite{Gopakumar:2005fx}
\be{schstrb}
\sigma_r={1\over \tau_r}= l_r = \int_{e_r}\sqrt{\phi(z,\e_i, p_a)}dz. 
\ee
We use this to make a change of variables of the Schwinger integrand 
from the $\sigma_r$ to the complex moduli $\e_i$ (as well as the 
residues $p_a$). Performing the integral over the $\{p_a\}\in R_+^n$ leaves us 
with an integral over ${\cal M}_{g,n}$ parametrised by the usual 
complex moduli $\e_i$. It is the integrand here that we can take to be a 
candidate correlator of the worldsheet CFT. 

\subsection{Strebel differentials for the four-punctured sphere}       

Since we are interested in planar four point functions 
(in the free theory, to begin with) we will look at the Strebel 
differentials on the four punctured sphere. 
This can always be put in the following form
\be{strebdif}
\phi(z)dz^2 = -C\frac{(z^2 -1)(z^2k^2 -1)}{(z-z_0)^2 (z-z_1)^2(z-z_2)^2
(z-z_3)^2} dz^2.
\ee
Here we have chosen the double poles to be at 
at $z_0, z_1, z_2, z_3$ and used 
the freedom of $SL(2, C)$ transformations to put the 
zeros at $\pm 1, \pm 1/k$.
Both $C$ and $k$ are as yet undetermined 
complex contants.
This parametrization of the 
Strebel differential assumes no additional symmetries. 
By Strebel's theorem, given the cross ratio $\e$ of the punctures and 
the residues at the four punctures, the constants $C$ and $k$ as well as the locations of the poles
$z_i$  are 
determined. Below we will see how this happens. 

\vspace{.5cm}
\noindent
{\emph{The $u$-plane}}
\vspace{.5cm}

From \eq{strebdif}, we see that $\int\sqrt{\phi(z)}dz$ would be 
most naturally expressed in terms of elliptic functions.  
Therefore we introduce the auxiliary 
$u$-plane where the doubly periodic properties 
of the differential $\sqrt{\phi(z)} dz$ are more manifest.
With $w^2 = (z^2-1)(z^2k^2 -1)$, we define the variable $u$ to be
\be{defu}
u = \int_{1}^z \frac{dz}{w}.
\ee
Upto a constant shift this is essentially the defining relation 
for the Jacobi elliptic 
function $\sn(u)$ of modulus $k$. We have the relation
\be{ivuz}
z = \sn(u + \frac{1}{2}\omega_1) = \frac{\cn(u)}{\dn(u)},
\ee
where $2\omega_1\equiv 4K(k)$ is one of the periods of $\sn(u)$ (with 
modulus $k$). 
Note that at $u=0$, $z=\sn(K)=1$
which fixes this integration constant. 
For future reference we note that the second period of the $\sn(u)$ 
function is given by $2\omega_2=2iK'(k)$. In other words,
the torus that the 
$u$-plane defines has periods $(2\omega_1, 2\omega_2)$. 
In appendix A. we gather together 
some basic facts (and notation) regarding elliptic functions 
which we will use.

We can now write the differential $\sqrt{\phi(z)}dz$ in the 
$u$-plane\footnote{By abuse of notation we will use the same symbol
$\phi$ for the $u$-plane as in the $z$-plane,
though the functional form of $\phi$ is different in both cases.} 
\be{stlup}
\sqrt{\phi(u)} du =  -i \sqrt{C}k^{\prime 4} \frac{\sn^2(u) }
{\prod_{i=0}^3 ( \cn(u) - z_i \dn(u) )}  du,
\ee
where $k'=\sqrt{1-k^2}$. We have used various shift identities
for the Jacobi functions \eq{shiftelip} and the formulae
for their derivatives \eq{derellipfn} to  arrive at this form.
{}From  \eq{stlup}, together with the locations of zeroes and poles 
of the  Jacobi  elliptic functions, we can see that 
$\sqrt{\phi(u)}$ has double zeros at 
$0, \omega_1, \omega_2, \omega_1 +\omega_2$ (modulo the 
double periodicity $(2\omega_1, 2\omega_2)$).
Furthermore from \eq{stlup} it is easy to see that both
$u_i$ and $-u_i$ (where $\cn(u_i)/\dn(u_i) = z_i$; $i=0\ldots 3$) are 
{\it simple} poles of $\sqrt{\phi(u)}$ with residues
\bea{residue}
r_i &=& -i\sqrt{C}k^{\prime 2}
\frac{\sn (u_i) \dn (u_i) }{\prod_{j\neq i}(\cn(u_i) - z_j \dn(u_i) )},\cr
&=& -i\sqrt{C}k^{\prime 2}\frac{\sn (u_i)\prod_{k=0}^3\dn (u_k)}
{\prod_{j\neq i}(\cn(u_i)\dn(u_j) - \cn(u_j)\dn(u_i))}.
\eea
Note that these properties on the $u$-plane (double zeroes and simple poles)
are not in contradiction 
with the different behaviour on the $z$-plane since this auxiliary torus is a 
branched cover of the original sphere. 

We would like to determine the positions of the poles $z_i$ 
(or $u_i$) and thus their cross ratio, in terms of the $r_i$.
Either by writing the Strebel differential as 
\be{strsum}
\sqrt{\phi(u)}du \propto \sum_{i=0}^3 r_i \frac{\sn (u_i)}{\dn (u_i)^2}
\frac{\dn(u)}{\cn(u) - z_i\dn(u)},
\ee
and demanding that it have the right double zeroes 
or else from direct verification one has the relations  
\bea{ficonzem}
\sum_i r_i \frac{1}{\sn(u_i)} = 0, \cr
\sum_i r_i \sn(u_i) =0, \cr
\sum_i r_i \frac{\cn(u_i) \dn(u_i)}{\sn(u_i)} =0.
\eea
In fact, if we substitute $r_i$ given in \eq{residue}
into \eq{ficonzem}  
then those
three equations are equivalent to the algebraic identities
\bea{consch}
\sum_i \frac{1}{\prod_{j\neq i} (z_i - z_j)}  =0, \cr
\sum_i \frac{z_i}{\prod_{j\neq i} (z_i - z_j)}  =0, \cr
\sum_i \frac{z_i^2}{\prod_{j\neq i} (z_i - z_j)}  =0, 
\eea
where $z_i = \cn(u_i)/dn(u_i)$. 
In principle the equations \eq{ficonzem} determine three
of the four $u_i$ in terms of the fourth, say $u_0$, as well as  the $r_i$. 
To determine $u_0$ we also need to know the Strebel lengths 
between zeroes.

\vspace{.5cm}
\noindent
{\emph{The Strebel lengths}}
\vspace{.5cm}

The main reason to go 
to the $u$-plane is that it enables us 
carry out the Strebel integrals between the zeroes.
The two independent lengths can be taken to be 
\be{strba}
a=\int_0^{\omega_1}\sqrt{\phi(u)}du
\ee
and 
\be{strbb}
b=- \int_0^{\omega_2}\sqrt{\phi(u)}du.
\ee 

To carry out these integrals it
will be useful to use an alternative representation of the 
Strebel differential in terms of the 
Weierstrass functions (see Appendix A.)   
\be{zfnpas}
\sqrt{\phi(u)}du = i \sum_i  r_i\left(
\zeta(u +u_i) - \zeta( u-u_i)  - 2 \zeta(u_i) \right)du.
\ee
By the properties of $\zeta(u)$, we can see that this has simple poles
at $u=\pm u_i$ with residue $\pm r_i$. It can also be verified 
(as shown in appendix B.) that this has the right double zeroes if 
the relations \eq{ficonzem} hold. From the uniqueness 
properties of elliptic functions
this is sufficient to conclude that \eq{zfnpas} is the same
as \eq{stlup}. 

Using the fact that $\zeta(u)$ is the derivative of a quasiperiodic 
function, we can easily evaluate the integrals in equations \eq{strba},\eq{strbb}
to obtain
\bea{fstbl}
a = \sum_i r_i \left[ \pi - 
2i ( \zeta(u_i) \omega_1 - \zeta(\omega_1) u_i ) \right] , \cr
b = \sum_i r_i \left[ \pi + 
2i ( \zeta(u_i) \omega_2 - \zeta(\omega_2) u_i ) \right] . 
\eea
For the future we will also record a useful linear combination 
of these two equations.  
\be{comstbl}
\pi \sum_i r_i u_i = (\pi \omega_1 + \pi \omega_2) \sum_i r_i 
- a\omega_2 - b\omega_1,
\ee
where we have used the  Legendre relation 
$2 (\omega_2 \zeta(\omega_1) -  \omega_2\zeta(\omega_2)) = i\pi$.

The strategy we will adopt is to assume one is given the 
residues $r_i$ and the Strebel lengths $a, b$ -- these are
after all linear combinations of Schwinger parameters. 
It is clear from \eq{ficonzem} we can eliminate all the poles
$u_1, u_2, u_3$ in terms of $u_0$. Then from the two
real equations \eq{fstbl} we can, in principle, determine the 
pole $u_0$ in terms of $a, b, r_i$. 
Thus the cross ratios of the poles $z_i={\cn{u_i}\over \dn{u_i}}$ is determined in terms of 
$r_i$ and $(a,b)$ as required. 
In practice, this is a difficult
task to carry out explicitly. 

\section{The Strebel differential for the $Y$ diagram}

Aharony et. al. \cite{Aharony:2006th}
made the nice observation that the Strebel 
differential corresponding to a $Y$ shaped skeleton Feynman diagram
is quite simple. Moreover, this simple differential
exists for every point on the 
moduli space of the four punctured sphere. 
In other words, as we vary the Schwinger parameters for the 
$Y$ diagram, we cover the entire moduli space and not just 
some subspace. In this section, we will look at this simple 
Strebel differential from the point of view of the general
framework outlined in the previous section. While this is not necessary 
for obtaining the final result \eq{etaso}, which can be obtained by simpler methods,
the general framework will be useful in developing a systematic perturbation 
expansion around this simple  differential. For completeness,
in appendix C., we will
outline the elementary way of obtaining the solution which is essentially
equivalent to the way in which it was solved in \cite{Aharony:2006th}.

The simplification of the $Y$ diagram is that the dual graph has three  
edges, one vertex and four faces. In fact, the valency of the vertex is six
corresponding to a fourth order zero for the differential, as per the 
general properties mentioned in section 2.1. Therefore instead of elliptic
functions, $\sqrt{\phi(z)}$ is algebraic and we can easily solve for the 
Strebel conditions \cite{Aharony:2006th} (See also appendix C.). 
The fourth order zero is clearly a 
limit of the separated zeroes in a generic Strebel differential. 
How do we take this limit in the general differential of
the form \eq{strebdif}? Notice that if the poles $z_i={z^{\prime}_i \over
\epsilon }$ with $z^{\prime}_i$ finite and $\epsilon \rightarrow 0$, then it 
would be natural to make the scaling of the variable $z={z'\over
\epsilon}$ in \eq{strebdif}. In other words we are taking a large $z$ 
limit. 
This effectively makes all zeroes coincide 
in the $z^{\prime}$ plane while keeping the poles finite. Note that $k$ is
not being scaled with $\epsilon$.

As mentioned in section 2., we 
use the equations \eq{ficonzem}, \eq{fstbl} and 
\eq{comstbl} to write the cross ratio $\e$ on the world sheet
in terms of the perimeters $r_i$ and the Strebel lengths $a, b$.
We now combine this with a systematic expansion scheme in $\epsilon$. 
The strategy of the approximation scheme is to first perform a Taylor
series expansion of \eq{ficonzem}, \eq{fstbl}, \eq{comstbl} 
about the specific point $\bar u$ in the $u$-plane, which corresponds to 
large $z$. Then
we can solve for $\e$ order by order in a
series expansion in $\epsilon$.

Solving for $u$ in terms of $z$,
using the change of variables in \eq{defu}, 
we can set up the following asymptotic expansion
\bea{uinz}
u &=& \int_1^\infty \frac{dz}{w} - 
\int_{\frac{z'}{\epsilon}}^\infty \frac{dz}{w}, \cr
&=& \bar{u} - \frac{\epsilon}{kz'} - 
\frac{1}{6k}(1 + \frac{1}{k^2}) \frac{\epsilon^3}{z^{'3}}
+ \cdots.
\eea
Here we have expanded the integrand in $\epsilon$ and  
integrated term by term, showing terms upto $O(\epsilon^3)$ for 
future purposes. 
In \eq{uinz}
$\bar u$ is given by
\bea{defbu}
\bar u &=& \int_1^\infty \frac{dz}{\sqrt{( z^2-1) (z^2k^2 -1)}},
\cr
&=&\frac{\omega_1}{2} + \omega_2.
\eea
The easiest way to fix $\bar u$ is to look at the equation for 
$z$ in terms of $\bar u$ for large $z$. From \eq{uinz} this is given by 
\be{pozu}
z\sim -\frac{1}{k (u - \bar u)}.
\ee
This indicates that $\bar u$ is a pole  with residue $-\frac{1}{k}$.
The fact that $z = \sn( u + \omega_1/2)$, 
and that $\sn(u)$ has a pole with residue $-1/k$ at $\omega_1 + \omega_2$
(see table 1)
fixes $\bar u = \omega_1/2 + \omega_2$.

To perform the large $z$ expansion in \eq{ficonzem} we need to write
$\sn(u_i), \cn(u_i),\dn(u_i)$ in terms of the asymptotic expansion in $1/z_i$
around the point $\bar u$.  This is done in appendix D, and from
\eq{simzeqnsd}, we can read off the leading terms which are relevant 
for us in this section. 
\be{orzeqnsld}
\sum_{i=0}^3 r_i = 0,
\qquad
\sum_{i=0}^3 r_i x_i = 0,
\qquad
\sum_{i=0}^3 r_ix_i^2 = 0.
\ee
Here $x_i\equiv {1\over z^{\prime}_i}$ are finite as 
$\epsilon \rightarrow 0$. 
As can be seen from the scaling of $\phi(z)$ with 
$\epsilon$ or as we will see explicitly in Sec.5, the equations \eq{fstbl} and
\eq{comstbl} for the Strebel lengths $a$ and $b$ can also be 
expanded in powers of $\epsilon$. The leading nonzero terms are of 
$O(\epsilon^3)$. So for the leading order solution of the Strebel 
conditions, we will not need these equations.

Therefore we have only the leading order equations in \eq{orzeqnsld} to solve.
In terms of $x_i=y_i+x_0$, we can easily verify that 
\eq{orzeqnsld} simplifies to 
\bea{leadeqn}
r_0 + r_1 + r_2 + r_3   
&=& 0,
\cr
r_1y_1 + r_2y_2 + r_3 y_3 
 &=&0,
\cr
r_1y_1^2 + r_2 y_2^2 + r_3y_3^2 
&=&0.
\eea
In other words, the translational mode $x_0$ drops out of these 
equations. 
Note that the first equation is a constraint among the perimeters (residues) of the
Strebel differential. 
The object of interest to us, the cross
ratio of the poles
\bea{defcr}
\eta &=& \frac{(z_3 -z_2)( z_1-z_0)}{(z_1-z_2)(z_3-z_0)},
\cr
&=& \frac{y_1( y_3 - y_2)}{y_3(y_1 -y_2)}.
\eea
depends only the ratio of the $y$'s (and is also independent of $x_0$).
In fact, defining 
\be{quadsol}
w_1 =\frac{y_1}{y_3}, \qquad
w_2 = \frac{y_2}{y_3},
\ee
the cross ratio  \eq{defcr} is simply
\be{defcrw}
\eta = w_1 \frac{1 -w_2}{w_1 -w_2},
\ee
We can solve for
$w_1, w_2$ using the last two equations of \eq{leadeqn}.
Eliminating $w_2$ in favour of $w_1$ using the second 
equation in \eq{leadeqn} and substituting it into the third equation 
yields
\be{quadeq}
r_1(r_1+ r_2) w_1^2 + 2 r_1r_3 w_1 + r_3( r_3 + r_2) =0.
\ee
We have 
\be{wonesol}
w_1 =\frac{y_1}{y_3} = \frac{ - r_1r_3 \pm 
\sqrt{r_1r_2r_3r_0}}{r_1(r_1 + r_2)},
\ee
and as a result 
\be{wtwosol}
w_2 = \frac{y_2}{y_3}= \frac{ -r_2r_3 \mp \sqrt{r_1r_2r_3r_0}}{r_2(r_1+r_2)}. 
\ee
To connect with the positive Strebel lengths, we will take 
$r_0=-p_0 <0$ and $r_i=p_i$ for $(i=1,2,3)$. This reflects the fact that 
$p_0=p_1+p_2+p_3$ is the relation among the perimeters for the dual graph to the 
$Y$ diagram.

Then, on substituting the values for $w_1$ and $w_2$ into \eq{defcrw} 
we obtain, after some simplifications
\be{etaso}
\e= \left( \frac{ \sqrt{p_0p_2} \pm i \sqrt{p_1p_3 }}
{p_1 + p_2} \right)^2.
\ee
Note that this 
zeroth order solution did not require knowledge of the 
modulus $k$. 
In appendix C. we obtain this result in an elementary fashion
and show how it is equivalent to the solution obtained in 
\cite{Aharony:2006th}. Without loss of generality, we can take 
the plus sign in \eq{etaso}, since the other choice is just the complex conjugate.

\section{A worldsheet four point function}

What we saw in the previous section is that for a $Y$ shaped 
skeleton Feynman diagram the Strebel differential is particularly
simple. Aharony et.al.  \cite{Aharony:2006th} 
exploited this property to study the contribution from the $Y$ diagram 
to the correlator $\la {\rm Tr}\Phi^{J_1}(x_1){\rm Tr}\Phi^{J_2}(x_2)
{\rm Tr}\Phi^{J_3}(x_3){\rm Tr}\Phi^{J}(0) \ra$ 
in free field theory $(J=J_1+J_2+J_3)$. 
As mentioned in the introduction, here we will instead 
study the correlator
\be{Ycorr}  
\Gamma^{(4)}_{ \{J_i\} }(x_1,x_2,x_3) = \la{\rm  Tr} X^{J_1}(x_1){\rm Tr }Y^{J_2}(x_2)
{\rm Tr }Z^{J_3}(x_3){\rm Tr}(\bar{X}^{J_1}\bar{Y}^{J_2}\bar{Z}^{J_3})(0)\ra,
\ee
since it gets contributions only from the $Y$ diagram (see footnote \ref{aharony}) in the free theory.
The answer in the free field theory is, of course, simply
\be{Yval}  
\la {\rm Tr} X^{J_1}(x_1){\rm Tr} Y^{J_2}(x_2)
{\rm Tr }Z^{J_3}(x_3){\rm Tr}(\bar{X}^{J_1}\bar{Y}^{J_2}\bar{Z}^{J_3})(0) \ra = {\tilde{C}(J_i)\over x_1^{2J_1}x_2^{2J_2}
x_3^{2J_3}}.
\ee
In the position space Schwinger representation, we can write this as 
\bea{Yschw}
&& \la {\rm Tr} X^{J_1}(x_1){\rm Tr} Y^{J_2}(x_2)
{\rm Tr} Z^{J_3}(x_3){\rm Tr}(\bar{X}^{J_1}\bar{Y}^{J_2}\bar{Z}^{J_3})(0) \ra \cr &=&
 C(J_i)\int_0^{\infty}d\s_1d\s_2d\s_3\s_1^{J_1-1}\s_2^{J_2-1}
\s_3^{J_3-1}e^{-(\s_1x_1^2+\s_2x_2^2+\s_3x_3^2)}.
\eea

We now need to change variables from the three $\s_i$ to $\e$ and 
an overall scaling factor, to be able to write \eq{Yval}
as a closed string integral. The three $\s_i$ are identified as per 
\cite{Gopakumar:2005fx} with the three independent Strebel lengths corresponding to 
the dual graph.  
They are, in fact, the three independent residues at the poles
$0,1$ and $\infty$ of the Strebel differential in \eq{strebdif}.
\be{sigp}
\s_i=p_i, \qquad  (i=1,2,3).
\ee 
We saw that the complex cross ratio $\e$ is given by \eq{etaso} to be 
\be{etagn}
\e= \left( \frac{ \sqrt{p_0p_2} + i \sqrt{p_1p_3 }}
{p_1 + p_2} \right)^2,
\ee
with $p_0=p_1+p_2+p_3$. Notice that $\e$ actually
depends only on the two independent
ratios $s_1={p_1\over p_3} ={\s_1 \over \s_3}$ and $s_2={p_2\over p_3} ={\s_2 \over \s_3}$
\be{etas}
\e= \left( \frac{ \sqrt{s_0s_2} + i \sqrt{s_1}}
{s_1 + s_2} \right)^2,
\ee
with $s_0={p_0 \over p_3}=1+s_1+s_2$.

We can solve \eq{etas} for $s_1$ and $s_2$ in terms of $\e$. Firstly, it is easy 
to check that with $\e=\me e^{i\theta}$, we have
\be{modeta}
\me={1+s_2\over s_1+s_2} ,~~~~~~~~~~~ 1-\cos{\theta}
={2s_1\over (1+s_2)(s_1+s_2)}.
\ee
Therefore
\be{modone}
\mone={1+s_1\over s_1+s_2},
\ee
so that one has
\be{sone}
s_1={(1-\me+\mone)\over (-1+\me +\mone)},
\ee
and 
\be{stwo}
s_2={(1+\me-\mone)\over (-1+\me +\mone)}.
\ee
Finally, 
\be{szero}
s_0= 1+s_1+s_2={(1+\me+\mone)\over (-1+\me +\mone)}.
\ee

Coming back to \eq{Yschw}, we can write it as
\bea{Yschwgn} 
\Gamma^{(4)}_{ \{J_i\} }(x_i) &=&
C(J_i)\int_0^{\infty}d\s_3\s_3^{J-1}
\int ds_1ds_2s_1^{J_1-1}s_2^{J_2-1}
e^{-\s_3(s_1x_1^2+s_2x_2^2+x_3^2)} \cr
&=& (J-1)!C(J_i)\int ds_1ds_2{s_1^{J_1-1}s_2^{J_2-1}\over
(s_1x_1^2+s_2x_2^2+x_3^2)^J}.
\eea
Note that $J=J_1+J_2+J_3$.
The measure term 
\be{meas}
ds_1ds_2=J(\e,\eb)d^2\e
\ee
where the Jacobian can be explicitly worked out, using \eq{sone}\eq{stwo}, to be
\be{jac}
J(\e,\eb)={1\over D^3}{|\e-\eb|\over \me \mone},
\ee
with $D=(-1+\me +\mone)$.
Using the identity
\be{etaid}
|\e-\eb|=\big[(1+\me +\mone)(1-\me+\mone)(1+\me-\mone)(-1+\me +\mone)\big]
^{1\over 2},
\ee
and equations \eq{sone},\eq{stwo}
we can finally rewrite \eq{Yschwgn} as
\bea{Yres}
\Gamma^{(4)}_{ \{J_i\} }(x_i)\equiv 
\int d^2\e G^{\{J_i\}}_{\{x_i\}}(\e, \eb)
= \int d^2\e{(1+\me +\mone)^{1\over 2}\over \me \mone}\times  \cr 
{(1-\me+\mone)^{J_1-{1\over 2}}(1+\me-\mone)^{J_2-{1\over 2}}
(-1+\me +\mone)^{J_3-{1\over 2}}\over [x_1^2(1-\me+\mone)+x_2^2(1+\me-\mone)
+x_3^2(-1+\me +\mone)]^J}.
\eea
We thus obtain the expression \eq{Yresult} for the integrand 
$G^{\{J_i\}}_{\{x_i\}}(\e, \eb)$  
on the moduli space of the four punctured sphere.

As mentioned in the introduction,  
for $G^{\{J_i\}}_{\{x_i\}}(\e, \eb)$ to be interpretable 
as a worldsheet
four point function  of physical vertex operators \eq{wscorr},
a number of requirements on its form have to be satisfied.
The simplest is that of crossing symmetry.
\bea{crossagn}
G^{J_2 J_1 J_3 J}_{x_2 x_1 x_3}(1-\e,1-\eb)&=&
G^{J_1 J_2 J_3 J}_{x_1 x_2 x_3}(\e,\eb),\cr
G^{J_3 J_2 J_1 J}_{x_3 x_2 x_1}({1\over \e},{1\over \eb})&=&
\me^4 G^{J_1 J_2 J_3 J}_{x_1 x_2 x_3}(\e,\eb).
\eea
The weight $\me^4$ in the second 
line is the same as the requirement that ${\cal V}^{J}$ in \eq{wscorr}
is a $(1,1)$ vertex operator
on the worldsheet. 
Note that the crossing symmetry requirement 
is not just a consequence of $SL(2,C)$ invariance, but also of the 
fact that there is an additional permutation symmetry among the labels
when we consider a correlator of primary operators. This is reflected 
in the fact that the functional form of the correlator $G^{\{J_i\}}_{\{x_i\}}$ is 
unchanged in Eqs. \eq{crossagn}. 

It is a nice
feature of the change of variables we have used that crossing 
symmetry is built in. We can indeed check that the change of variables
in equations \eq{sone}\eq{stwo} implements the permutation 
symmetry between the labels $(1,2,3)$ in the way we expect\footnote{We 
thank A. Sen for useful discussions on this point.}. Therefore it is 
assured that the $G^{\{J_i\}}_{\{x_i\}}(\e, \eb)$ will satisfy the
requirements of \eq{crossagn}. This is, in fact, 
easy to verify explicitly from the expression \eq{Yresult}. 

Another requirement for a CFT correlator in a string theory is having an
OPE consistent with locality. In other words, in the expansion \eq{ope} in powers
$\e^h\eb^{\bar h}$, the correlator must have $(h-\bar{h})$ always an integer.
In \cite{Aharony:2006th}
the form of the correlator was given 
to the first couple of orders in a somewhat different variable $t$ 
(which is to leading order the same as $\e$)
and it was found that the powers obeyed this property, though $h,\bar{h}$ could
be individually half integers. 
 From the explicit expression in \eq{Yres},  
this is actually manifest to {\it all orders} in the expansion for the 
$G^{\{J_i\}}_{\{x_i\}}(\e, \eb)$.  

Finally, we would like to be able to see if $G^{\{J_i\}}_{\{x_i\}}(\e, \eb)$
actually arises as the correlator of {\it local} vertex operators. 
The fairly simple dependence of $G^{\{J_i\}}_{\{x_i\}}(\e, \eb)$
on $\me$ and $\mone$ are indicative of local dependence on the 
points $0, 1, \infty$ and $\e$. Indeed, as was pointed out to us by 
D. Gaiotto, the expressions that enter here are very closely related
to similar ones in the four point functions of spin fields in the 
Ising model. 

For instance, the correlator of four order operators $\s(z,\bar{z})$
(see equation (12.63) of \cite{DiFrancesco:1997nk}) is given by
\be{fourord}
\la \s(1)\s(2)\s(3)\s(4)\ra =\sqrt{1\over 2|z_{13}z_{14}|^{1\over 2}}
{1\over \sqrt{\me\mone}}(1+\me +\mone)^{1\over 2}.
\ee
and that of two order and two disorder 
operators ( equation (12.66) in \cite{DiFrancesco:1997nk})
is
\be{twoord}
\la \s(1)\mu(2)\s(3)\mu(4)\ra =\sqrt{1\over 2|z_{13}z_{14}|^{1\over 2}}
{1\over \sqrt{\me\mone}}(-1+\me +\mone)^{1\over 2}.
\ee 
By permuting the ordering of the operators in \eq{twoord}, we get 
the last factor proportional in turn to $(1-\me+\mone)^{1\over 2}$
and $(1+\me-\mone)^{1\over 2}$. 
These are exactly the ingredients that enter into the correlator 
defined by \eq{Yres}. Thus it is not far fetched to imagine that 
the entire expression in \eq{Yres} might arise as a local correlator
in a CFT. 
Since the spin field correlators seem like building blocks for
the actual object that appears here,
It would be very interesting to use them to try and reconstruct
the worldsheet CFT, at least in an algebraic way.
For instance, knowing the conformal blocks of the Ising model, we can try
and decompose our candidate correlator into candidate conformal blocks. 

Another potentially interesting connection suggested by the appearance of 
Ising model correlation functions is to $W$-strings (see
\cite{Hull:1993kf}\cite{West:1993np}\cite{Pope:1993qu} for
reviews). Ising model correlation function are known to appear
in the four point scattering amplitudes of $W_3$ strings \cite{Freeman:1992mb}\footnote{We thank 
Ari Pakman for bringing this reference to our attention.}
\footnote{The Ising model has also
arisen in $Spin(7)$ compactifications \cite{Shatashvili:1994zw}.}. 
 This maybe a natural connection 
given the fact that the free gauge theory has higher spin symmetries in {\it spacetime} which 
might well be reflected as higher spin symmetries on the {\it worldsheet} in the usual manner by which spacetime and worldsheet symmetries are related.

In any case, we believe there is enough reason to take $G^{\{J_i\}}_{\{x_i\}}(\e, \eb)$
seriously as a candidate four 
point function in the as yet unknown worldsheet theory of 
$AdS_5\times S^5$ dual to free ${\cal N}=4$ Yang-Mills theory.

\section{A perturbation scheme around the $Y$ diagram}

In this section we develop a systematic approximation scheme 
around the Y diagram. We use the equations \eq{ficonzem}, \eq{fstbl} and
\eq{comstbl} to write the cross ratio on the world sheet in terms of the 
perimeters $r_i$ and the Strebel lengths $a,b$. As we have seen in 
Section 3, the Y diagram corresponds to the large $z$ limit of the
general Strebel differential given in \eq{strebdif}. In the $u$-plane,
this point corresponds to $\bar u = \omega_1/2 + \omega_2$. 
In Section 3. we have already used the leading terms in the expansion 
to obtain the solution shown in \eq{etaso}.  
We now retain the first two 
leading terms in the equations \eq{simzeqnsd} and  the Strebel length
equations \eq{fstbl}, \eq{comstbl}  to obtain the leading correction 
in $\epsilon$ to the 
cross-ratio. From \eq{simzeqnsd} 
we see that the three conditions on the poles \eq{ficonzem} 
in this approximation reduce to 
\bea{orzeqns}
\sum_{i=0}^3 r_i - \epsilon^4 \tilde E \sum_{i=0}^3 r_i x_i^4 &=& 0,
\cr
\sum_{i=0}^3 r_ix_i - \epsilon^2 E \sum _{i=0}^3 r_i x_i^3 &=& 0,
\cr
\sum_{i=0}^3 r_i x_i^2 - \epsilon^2 E \sum_{i=0}^3 r_i x_i^4 &=& 0.
\eea
where 
\be{defc34}
\tilde{E}= - \frac{(1-k^2)^2}{ 8 k^4}, \qquad  
E = - \frac{1}{2} ( 1 + \frac{1}{k^2}).
\ee
It is clear from the expansions that we have
retained  terms only to  $O(\epsilon^4)$.
 
The next set of  equations to expand are \eq{fstbl} and
\eq{comstbl}.
We expand the variables $u_i$ occurring in these equations about $\bar u$ in 
terms of $x_i$ and performing simple algebraic manipulations (see
appendix D. for details)
we obtain the following equivalent equations valid up to $O(\epsilon^4)$
\bea{eqfab}
a &=& \epsilon^3 p_1 \sum r_i x_i^3 + \epsilon^4 p_2 \sum r_i x_i^4, \cr
a\omega_2 + b\omega_1 
&=& \epsilon^3 q_1 \sum r_i x_i^3 + \epsilon^4 q_2\sum r_i x_i^4, 
\eea
where $p_1$, $q_1$, $p_2$, $q_2$ are constants depending on the 
value of $\bar u$, as given in \eq{defas}, \eq{cstblexp}\eq{defaabb}. 
For our purpose here, it is sufficient 
to provide the 
following information about the constants which are valid to the 
$O(\epsilon^0)$
\bea{inp1q1}
\frac{p_1^{(0)}}{q_1^{(0)}} &=&
2i\left( \frac{1-10k^2+ k^4 }{12}  \vartheta_3^2(q) + \frac{1}{12} 
\frac{1}{\vartheta_3^2(q) }\frac{\vartheta_1'''(q)}{\vartheta_1'(q)} 
\right),
\cr
q_1^{(0)} &=& -\frac{\pi}{3k}( 1 + \frac{1}{k^2}), \cr
\omega_1 &=& \pi \vartheta_3^2(q), \qquad q 
= \exp( 2\pi i \frac{\omega_1}{\omega_2 }).
\eea
The periods $(2\omega_1, 2\omega_2)$ of the auxilliary torus in the 
$u$ plane are functions
of the modulus $k$, they are all $O(\epsilon^0)$. 
In the above equation the superscript ${}^{(0)}$ refer to the order in $\epsilon$. 
It is clear from \eq{eqfab} that the Strebel lengths $a$ and $b$ begin 
at $O(\epsilon^3)$. 
This fact can also be seen easily by performing the scaling $z=z'/\epsilon$ in 
the equation for the Strebel differential \eq{strebdif}.
We can therefore expand the Strebel lengths as 
\be{expstrbep}
a = a^{(3)} \epsilon^3 + a^{(4)} \epsilon^4+ \ldots, 
\qquad
b = b^{(3)} \epsilon^3 + b^{(4)} \epsilon^4 + \dots.
\ee
For convenience we define
\be{defcpr}
c= \sum r_i x_i^3 , \qquad \tilde{c} = \sum r_i x_i^4 ,  
\ee
From \eq{eqfab} it is clear that $c, c'$ begins at $O(\epsilon^0)$. 
Now comparing terms of order $O(\epsilon^3)$ in the first equation of \eq{eqfab} we 
obtain the following relation
\be{solc10}
c^{(0)} =  \sum r_i x_i^{(0)3} = \frac{a^{(3)}}{p_1^{(0)}}.
\ee
Furthermore, comparing the terms of $O(\epsilon^3)$ in both the equations 
and eliminating $c$ from them we obtain 
\be{keq}
\frac{p_1^{(0)}}{q_1^{(0)}}
(a^{(3)} \omega_2^{(0)} + b^{(3)} \omega_1^{(0)})  =  
a^{(3)}.
\ee
It is this equation which determines $k$ in terms of $a^{(3)}$ and $b^{(3)}$.
One substitutes the values of $\omega_1$ and $\omega_2$ in terms of $k$, uses \eq{inp1q1} and
then solves for $k$ using the above equation. This is done in appendix E.
From now on, we assume that $k=k(a, b)$, and similarly
$\omega_1=\omega_1(a,b)$ and $\omega_2=\omega_2(a,b)$ since each of them are functions of
$k$. Note that $k$ starts at $O(\epsilon^0)$.

Now that we have the basic equations \eq{orzeqns}, \eq{eqfab} and \eq{keq}
we can organize them so that one can solve for the $x_i$ using 
perturbation theory.
To isolate the translational degree of freedom in the variables $x_i$ we 
first write $x_1 = y_1 + x_0, x_2 = y_2 + x_0, x_3 = y_3+ x_0$.
Therefore the variables to solve for are now $y_1, y_2, y_3, x_0$. 
Then the equations \eq{orzeqns} and \eq{defcpr}  reduce to 
\bea{petyeq}
\sum_{i=1}^3 r_i &=& \epsilon^4 \tilde{E} \tilde c, \cr
\sum_{i=1}^3  r_i y_i  + \epsilon^4 \tilde E \tilde c x_0 &=& \epsilon^2 E c, \cr
\sum_{i=1}^3 r_i y_i^2 
+ 2  \epsilon^2 E c  x_0 
- \epsilon^4 \tilde E \tilde c x_0^2 &=& \epsilon^2 E \tilde c,
\cr
\sum_{i=1}^3 r_i y_i^3
+ 3  \epsilon^2 E \tilde c x_0
- 3  \epsilon^2 E cx_0^2
 + \epsilon^4 \tilde E \tilde c  x_0^3 &=& c,
\cr
\sum_{i=1}^3 r_i y_i^4
+ 4 c x_0 -6 \epsilon^2 E\tilde c x_0^2 
+4 \epsilon^2 E c x_0^3 
- \epsilon^4 \tilde E\tilde  c x_0^4 &=& \tilde c.
\eea
To obtain these equations we  have repeatedly used the basic equations
\eq{orzeqns} and \eq{defcpr}. Note that the first equation in \eq{petyeq} 
determines the constant $\tilde c $ since the perimeters $r_i$ are given to us.
The first equation indicates the constraint $\sum_{i=0}^3 r_i =0$ holds
till $O(\epsilon^4)$.
The constant $c$ is determined 
from \eq{solc10} and finally the constants $E$ and $\tilde E$ are defined in 
\eq{defc34}.  As mentioned earlier, 
the modulus $k$ is determined as a function of $a, b$ from 
\eq{keq}.  To summarize, \eq{petyeq} can be solved using 
perturbation theory for the variables $y_1, y_2, y_3, x_0$  in 
terms of the perimeters $r_i$ and the Strebel lengths $a, b$. 
We can obtain   the cross ratio from the equation \eq{defcrw}.

We now set about to solve the above equations
and obtaining the leading correction. 
From the first two equations in \eq{petyeq} we see that the next non-trivial correction
to the cross ratio occurs at  $O(\epsilon^2)$. 
These equations determine the ratios $w_1 = y_1/y_3, w_2 =y_2/y_3$
to $O(\epsilon^2)$ 
\bea{rateq2}
r_1 w_1 + r_2 w_2 + r_3 &=& \epsilon^2 \frac{Ec^{(0)}}{y_3^{(0)}}, \cr
r_1w_1^2 + r_2w_2^2 + r_3 + 2\epsilon^2 Ec^{(0)} \frac{x_0^{(0)}}{y_3^{(0)2}} &=& 
\epsilon^2 \frac{E\tilde c^{(0)}}{y_3^{(0)2}}.
\eea
These equations indicate that we need to need the information of $y_3$ 
and $x_0$ 
at the zeroth order. This is obtained from 
using the last two equations of \eq{petyeq} at the zeroth order, these 
are
\bea{zy3x0}
\sum_{i=1}^3 r_i y_i^3 = c^{(0)}, \qquad
\sum_{i=1}^3 r_i y_i^4 + 4 c^{(0)} x_0 = \tilde c^{(0)}.
\eea
Substituting the zeroth order values of $y_1, y_2$ in terms of $r_i$ and 
$y_3$  in the above equation we obtain 
\be{soly3}
y_3^{(0)} = \left(  \frac{ c^{(0)} r_1 r_2(r_1 +r_2)^2 }
{ r_3 \{ r_1 r_2 r_3 (r_0 -r_3) \pm (r_2 -r_1) \sqrt{r_1r_2r_3r_0} \} }
\right ) ^{1/3},
\ee
where $c^{(0)}$ is obtained from \eq{solc10}.
Similarly 
from the second equation in \eq{zy3x0} we can obtain 
$x_0$ as 
\bea{solx0}
x_0^{(0)} &=& 
\frac{\tilde c^{(0)}}{4 c^{(0)}}
- \frac{ (y_3^{(0})^4}{4 c^{(0)}}
 \left( \frac{r_3r_0}{r_1r_2(r_1 + r_2)^3} \right) \times \\ \nonumber
&\;& \left\{ r_1 r_2 ( r_0r_3 - r_0^2 - 4r_3^2 ) + r_3r_0 ( r_1 + r_2)^2 
 \pm 4 r_3( r_2 -r_1) \sqrt{ r_1 r_2 r_3 r_0}
\right\}.
\eea
Here one has to substitute $y_3$ from \eq{soly3}, 
$c$ from \eq{solc10} and $\tilde c$ from the first line
of \eq{petyeq} 
\be{valctc}
 \tilde c^{(0)} = \sum r_i^{(4)}/\tilde E.
 \ee
Now that we have all the quantities entering into \eq{rateq2},
we can solve for $w_1$ and $w_2$.  For convenience we rewrite \eq{rateq2} as
\bea{eqforw}
r_1 w_1 + r_2 w_2 &=& - r_3 - 
\epsilon^2\delta_1,  \cr
r_1 w_1^2 + r_2 w_2^2 &=& -r_3 -
\epsilon^2 \delta_2,
\eea
where 
\bea{defdelta}
\delta_1 &=& - \frac{E c^{(0)} }{y_3^{(0)} }, \qquad
\delta_2 = - 
\frac{E}{y_3^{(0)2 }}  (  \tilde c^{(0)} - 2 c^{(0)} x_0^{(0)} ) .
\eea
Eliminating $w_2$ using the first equation of \eq{eqforw} in 
the second equation we obtain the following quadriatic equation
where we have retained terms to $O(\epsilon^2)$.
\be{quadweq}
r_1( r_1+ r_2) w_1^2 + 
( 2 r_1 r_3 + 2 \epsilon^2 r_1 \delta_1) w_1 
+ r_3( r_3 + r_2) + r_2 \epsilon^2 \delta_2  + 2 r_3 \epsilon^2 \delta_1 =0.
\ee
The solution for $w_1$ and $w_1$ are given by
\bea{solw1w2}
w_1 &=&- \frac{1}{r_1(r_1+ r_2)} \left[  r_1 r_3 \pm \sqrt{r_1 r_2r_3r_0}\right. \cr
&+& \left. \epsilon^2\left( r_1\delta_1 \pm 
\sqrt{r_1r_2r_3r_0} ( -\frac{ \delta_1}{  r_0} + ( \frac{1}{r_3} + \frac{1}{r_0} )
\frac{\delta_2}{2} ) \right) \right],
 \cr
w_2 &=& - \frac{1}{r_2( r_1+r_2)} \left[  r_2 r_3  \mp \sqrt{r_1 r_2 r_3 r_0}\right. \cr
&+&\left. \epsilon^2\left(  r_2\delta_1 
\mp \sqrt{r_1r_2r_3r_0} ( -\frac{ \delta_1}{  r_0} + ( \frac{1}{r_3} + \frac{1}{r_0} )
\frac{\delta_2}{2} ) \right)\right].
\eea
From this we can evaluate the cross-ratio using \eq{defcrw}, which is given by
\bea{corcross}
\eta_{\pm} &= &\eta^{(0)}_\pm 
+ \epsilon^2 \frac{  \sqrt{p_2p_0} \pm i \sqrt{p_1p_3} }{ (p_1 + p_2)^2 }
\left( \delta_1 ( \sqrt{\frac{p_2}{p_0} }\pm i \sqrt{\frac{p_1}{p_3} })
+ \kappa ( \sqrt{p_2p_0} \pm i \sqrt{p_1p_3} ) \right) \cr
&-& \epsilon^2 \kappa \eta^{(0)}_\pm, \cr
{\rm where}\,  \kappa &=& - \frac{\delta_1}{ r_0}  + (\frac{1}{r_3} + \frac{1}{r_0} )\frac{\delta_2}{2} .  
\eea
Here $\eta_\pm^{(0)}$ is the zeroth order solution for the cross ratio
given in \eq{etaso}.

\section{Conclusions}

We have seen that we can arrive at explicit expressions for candidate worldsheet correlators, 
in the string dual to free Yang-Mills theory by using a definite change  of variables on
the Schwinger representation of field theory amplitudes. These candidates seem quite promising 
in that they obey many of the properties one might expect of them. Moreover, their form seems to
suggest ways of potentially gaining a better understanding of the worldsheet theory. The relation
with the spin field correlators might be a fruitful way to both uncover the structure of conformal blocks
and maybe even give some clues about the worldsheet action. One point that needs better understanding here is the fact that the spacetime special conformal transformations do not 
act locally in this Schwinger representation \cite{Aharony:2006th}. Perhaps for the cases
of spacetime conformal invariance, we can modify the prescription to make this symmetry
manifest on the worldsheet as well.

It will also be important to have at hand a large number of explicitly worked out correlators. 
The case of extremal correlators is particularly interesting, affording a way to potentially compare with 
large radius intuition as well as methods. There are also 
other special four point diagrams which, as observed in 
\cite{Aharony:2006th}, seem to get contributions only from real subspaces of moduli space.
It would be good to get a better grasp of what is going on here by comparing with cases 
such as the $Y$ diagram which yield local expressions on the worldsheet.  

To summarise, we are hopeful that
this approach to obtaining the worldsheet theory will be fruitful, like in the Liouville
case, in effectively bypassing an action formulation of the theory and yet capturing 
all the essential information. 

\acknowledgments

We would like to thank A. Dhar, E. Gava, S. Giddings, D. Ghoshal, D. Gross, 
L. Motl, K.S. Narain, A. Pakman, J. Polchinski, A. Strominger and C. Vafa  
for useful discussions.
Special thanks are due to D. Gaiotto, S. Minwalla and A. Sen for insightful remarks during the course of this work. 
It is also a pleasure to thank O. Aharony, Z. Komargodski and S. Razamat for 
discussions and specially for their very useful comments on a preliminary draft
of this paper.  J.R.D. wishes to thank the Weizmann Institute,  ASICTP and CERN
for hospitality during the course of this work. R. G. would like to thank the string theory
group at Harvard University, the organisers of the "New Topological Structures in Physics" workshop at MSRI, Berkeley, the Physics Department, UC Santa Barbara and TIFR, Mumbai 
for hospitality while this work was being 
completed. We would also jointly like to thank the organisers  of the Indo-Israeli
workshop on String Theory at Ein Boqeq for their hospitality. Finally, we are indebted to 
the people of India for their unstinting support for our discipline.

\appendix
\section{Elliptic Functions}

In this section we briefly review the properties of elliptic functions which
are used  extensively in this paper. A detailed discussion is available in 
\cite{Watson}

\vspace{.5cm}
\noindent
{\emph{The Jacobian elliptic functions}
\vspace{.5cm}

The elliptic function $\sn u$ is defined as follows, consider the integral
\be{defbelipsn}
u = \int_0^z dt ( 1-t^2)^{-\frac{1}{2} } ( 1 - k^2 t^2)^{-\frac{1}{2}},
\ee
then $ z= \sn(u, k)$, we will supress the modulus $k$ unless different.
The complementary modulus $k'$ is defined by $k^2 + k^{\prime 2} =1$
The elliptic functions $\cn(u)$ and $\dn(u)$ are defined from
\be{sqrelelip}
\sn^2 u + \cn^2 u = 1, \qquad
k^2 \sn^2 u + \dn^2 u = 1.
\ee
The periods $K$ and $iK'$ of the elliptic functions are defined by the 
integrals
\be{defK}
K = \int_0^1 dt ( 1 - t^2 )^{-\frac{1}{2}} ( 1 - k^2 t^2)^{-\frac{1}{2}} ,
\ee
where the integral is along the straight line from $0$ to $1$. $K'$ is 
given by
\be{defKKp}
K' = \int_1^{1/k} dt (t^2 -1) ^{-\frac{1}{2}} ( 1- k^2 t^2)^{-\frac{1}{2}} ,
\ee
here the branch cuts are chosen from $0$ to $-\infty$ and $1$ to $\infty$. 
We define the periods $\omega_1$ and $\omega_2$ by 
\be{defperiod}
\omega_1 =  2K, \qquad\quad   \omega_2 = i K'.
\ee
Note that $K, K'$ and therefore $\omega_1, \omega_2$ are 
functions of the modulus $k$ through the integrals \eq{defK} and \eq{defKKp}.
From \eq{defbelipsn} it is easy to see that $\sn u$ is an odd function of $u$.
The elliptic functions $\cn u$ and $\dn u$ are even functions of $u$. 
In the table below we summarize the information about the periods, 
the poles and the respective residues and the zeros of the
corresponding elliptic functions. 

\vspace{.3cm}
\begin{center}
\begin{tabular}{l|l|l|l|l}
Function & Periods & Poles & Residues & Zeros \\
\hline
$\;$ & $\;$ & $\;$ & $\;$ & $\;$ \\
$\sn u$ & $(2\omega_1, 2\omega_2)$ & $\omega_2$& $\frac{1}{k}$ & $0$ \\
 $\;$ & $\;$  & $\omega_1 + \omega_2$  &$-\frac{1}{k}$ & $\omega_1$ \\
 $\;$ & $\;$ & $\;$ & $\;$ & $\;$ \\
 \hline
  $\;$ & $\;$ & $\;$ & $\;$ & $\;$ \\
    $\cn u $ & $(2\omega_1, \omega_1 + 2\omega_2)$ & $\omega_2$ &$-\frac{i}{k}$ & $\frac{\omega_1}{2}$ \\
 $\;$  & $\;$ & $ \omega_1+\omega_2$ & $\frac{i}{k}$ & $\frac{3\omega_1}{2}$ \\
$\;$ & $\;$ & $\;$ & $\;$ & $\;$ \\
\hline
$\;$ & $\;$ & $\;$ & $\;$ & $\;$ \\  
  $\dn u$ & $(\omega_1, 4 \omega_2)$ & $\omega_2$ & $-i$& $\frac{\omega_1}{2} + \omega_2 $ \\ 
 $\;$  & $\;$ & $3 \omega_2$ & $i $& $\frac{\omega_1}{2} + 3 \omega_2$
\\
$\;$ & $\;$ & $\;$ & $\;$ & $\;$ \\
\hline
\end{tabular}
\vspace{.5cm}
\\
{\bf\small{Table 1:}} General description of the Jacobian elliptic functions.
\end{center}
\noindent
The elliptic functions have the following shift properties. 
\be{shiftelip}
\sn(u + \frac{\omega_1}{2} ) = \frac{\cn u}{\dn u}, \quad
\cn( u + \frac{\omega_1}{2} ) = - k' \frac{\sn u}{\dn u}, \quad
\dn( u + \frac{\omega_1}{2} ) = k' \frac{1}{\dn u}.
\ee
We also need the following shift properties of the Jacobian elliptic functions
\bea{shiftellip1}
 \sn(u+\omega_1) = -\sn(u), \quad \cn(u+ \omega_1) &= &-\cn(u), \quad
\dn(u+ \omega_1) = \dn (u), \cr
 \sn(u +\omega_2) = \frac{1}{k \sn(u)},  \quad
\cn(u+ \omega_2) &=& -i\frac{\dn(u)}{k\sn(u) }, \quad
\dn(u+\omega_2) = -i \frac{\cn (u)}{ \sn(u)}, \cr
  \sn( u+ \omega_3) = - \frac{1}{k \sn(u)}, \quad
\cn( u + \omega_3) &=& i \frac{\dn(u)}{k\sn(u)}, 
\dn(u + \omega_3) = -i \frac{\cn(u)}{\sn(u)}.
\eea
where $\omega_3 = \omega_1 + \omega_2$.  
The derivatives of the elliptic functions are given by
\bea{derellipfn}
\frac{d}{du} \sn u &=& \cn u \dn u, \cr
\frac{d}{du} \cn u &=& -\sn u \dn u, \cr
\frac{d}{du} \dn u &=& - k^2 \sn u \cn u.
\eea

\vspace{.5cm}
\noindent
{\emph{ The Weierstrass $\wp$ function and related functions}}
\vspace{.5cm}

The Weierstrass function can be defined by the following expansion
\be{defwpfn}
\wp (z) = \frac{1}{z^2} + \sum_{m, n \neq 0} 
\left( \frac{1}{(z - 2 m\omega_1 - 2n \omega_2)^2} 
- \frac{1}{( 2 m\omega_1 - 2n \omega_2)^2} \right).
\ee
It is an even function with periods $(2\omega_1, 2\omega_2)$. The derivatives 
of the $\wp$ function vanish at the following points
\be{vanishder}
\wp'(\omega_1) =0, \quad \wp'(\omega_2) =0, \quad \wp'(\omega_1 + \omega_2) =0.
\ee
The Weierstrass $\wp$ function can be thought of as building blocks of 
elliptic functions since any elliptic function of period 
$(2\omega_1, 2\omega_2)$ can be expresses in terms of the $\wp$ function
of the same period. The expression is rational in $\wp(z)$ and linear in $\wp'(z)$.

Related to the Weierstrass function is the function $\zeta (z)$ which is defined as
\be{defzetafn}
\frac{d\zeta(z)}{dz} = - \wp(z).
\ee
The $\zeta(z)$ function is an odd function of $z$, it is quasi-periodic in the 
periods $\omega_1$ and $\omega_2$ with the quasi periods defined as
\be{quasipe}
\zeta(z+ 2\omega_1) = \zeta(z) + 2\eta_1, \qquad
\zeta(z+ 2\omega_2) = \zeta(z) + 2\eta_2,
\ee
where $\eta_1$ and $\eta_2$ are given by
\be{defetas12}
\eta_1 = \zeta(\omega_1), \qquad \eta_2= \zeta(\omega_2).
\ee
They satisfy the Legendre relation
\be{legrela}
2( \eta_1\omega_2 - \eta_2\omega_1 ) = \pi i 
\ee
Since the $\zeta(z)$ function is the integral of the $\wp$ function it is clear that
it has a simple pole at the origin $z=0$ with residue $1$.  
Finally we define the function $\sigma(z)$ by the equation
\be{defsigma}
\frac{d}{dz} \log  \sigma(z) = \zeta(z).
\ee
The $\sigma(z)$ functions is an odd function in $z$, it
 is useful for integration of the $\zeta (z)$ function and we
use it to perform the Strebel integrations. It satisfies the following quasi-periodic
relations
\be{quasisigma}
\sigma(z+ 2\omega_1) = - e^{2\eta_1( z + \omega_1)} \sigma(z), \qquad
\sigma(z+ 2\omega_2) = - e^{2\eta_2( z + \omega_2)} \sigma(z).
\ee

\vspace{.5cm}
\noindent
{\emph {Relationship between elliptic functions and Weierstrass ${\wp}$ function}}
\vspace{.5cm}

In this paper we extensively use the relationship between the Weierstrass
$\wp$ function and the Jacobian elliptic function $\sn(u)$. As mentioned
earlier, any  doubly periodic function of periods $2 \omega_1$ and $2\omega_2$
can always be written in terms of the Weierstrass $\wp$-function
having the same periodicity.
This is given by
\be{snpfn}
\sn(u + \frac{\omega_1}{2}) = 1 
+ \frac{1}{2} \frac{k^2-1}{(\wp (u) - \frac{5k^2-1}{12} )}.
\ee
Here we have used the theorm of uniformisation of curves of 
genus unity given in \cite{Watson}, page 454.   The above equation 
is easily  verified for $u=0$. 
There is another relation one can obtain using the $\wp$ function and the
Jacobian elliptic function. Let the values of the $\wp$ at half the periods be 
given by
\be{defe}
\p(\omega_1) = e_1, \qquad
\p(\omega_2) = e_2, \qquad
\p(\omega_3) = e_3. 
\ee
where $\omega_3 =\omega_1 +\omega_2$.
Then from \eq{snpfn} it can be shown that
\be{vale}
e_1 = \frac{1+k^2}{6}, \qquad
e_2 =- \frac{1 -6k + k^2}{12}, \qquad
e_3 = - \frac{1+6k + k^2}{12}.
\ee
We then have the following relation between the $\p$ function and 
the $\sn$ function.
\be{nerpsn}
\p(u ) = e_3 + \frac{e_1 -e_3}{\sn^2 (u\sqrt{e_1 -e_3} , \tilde k)}.
\ee
Here note that the $\p$ function is doubly periodic with periods 
$(2\omega_1,  2\omega_2)$, and the modulus of the 
$\sn(u)$ function is given by 
\be{nemod}
\tilde k = \sqrt{\frac{e_2 -e_3}{e_1-e_3} }= 2 \frac{\sqrt{k}}{1+k}.
\ee

\section{Strebel differential in terms of Weierstrass functions}

In this section we will write the general Strebel differential given in \eq{strebdif}
in two convenient forms. One involves Weierstrass $\wp$ functions and the
second involves the $\zeta(u)$ function. The former is useful to obtain 
relations between  the poles of the Strebel differential and the perimeters
while the latter is useful to perform the Strebel integration in \eq{strba} and \eq{strbb}.

\vspace{.5cm}
\noindent
{\emph{$\sqrt{\phi(u)}$ in terms of $\p$-function}}
\vspace{.5cm}

It is useful to write the differential $\sqrt{\phi(u)} du$ in terms
of the Weierstrass ${\wp}$ function for the purpose of obtaining the
Strebel lengths by integration as well as to obtain relations between the 
poles of the Strebel differential and the perimeters.
To do this we take the view that we are given the residues $r_i$ at 
the poles $u_i$ of the differential $\sqrt{\phi(u)}$ in \eq{stlup}.
It is clear from 
the transformation between the $z$-plane to the $u$-plane given in \eq{ivuz},
that the differential $\sqrt{\phi(u)}$ is a doubly periodic function of
periods $(2\omega_1, 2\omega_2)$.
Then the  unique doubly periodic function of periods $(2\omega_1, 2\omega_2)$
with the residues $\pm r_i$ at $\pm u_i$ and a double zero at $u=0$ is given by

\be{pfnpar}
\sqrt{\phi(u)} = i
\sum_{i =0}^{3} r_i \frac{\p'(u_i)}{\p(u) - \p(u_i)} .
\ee
We also have that the above function must have double zeros 
at $\omega_1, \omega_2$ and $\omega_1+\omega_2$. 
Demanding that the function in \eq{pfnpar} has zeros
at these points  gives the following
conditions
\bea{zerocond}
\sum_{i=0}^3 r_i \frac{ \p'(u_i)}{\p(\omega_1) - \p(u_i) }  =0,
\cr
\sum_{i=0}^3 r_i \frac{ \p'(u_i)}{\p(\omega_2) - \p(u_i) }  =0,
\cr
\sum_{i=0}^3 r_i \frac{ \p'(u_i)}{\p(\omega_3) - \p(u_i) }  =0.
\eea
where $\omega_3 = \omega_1 +\omega_2$. 
It is clear that the derivative of the function $\sqrt{\phi(u)}$ 
given in \eq{pfnpar}
at the points $\omega_1, \omega_2, \omega_3$ vanishes since
$\p'(\omega_i) =0$. The equations \eq{zerocond} together with the 
fact that the derivative of the function $\sqrt{\phi(u)}$ vanishes
at the points $\omega_1, \omega_2, \omega_3$ ensures that
the representation of the Strebel differential in terms of 
the $\wp$ function in \eq{pfnpar} has double zeros at these points.
The  three equations in \eq{zerocond} 
enables one to determine the position of the poles $u_1, u_2, u_3$. 
in terms of $u_0$ and the residues $r_i$. 

\vspace{.5cm}
\noindent
{\emph{Conditions on zeros}}
\vspace{.5cm}

Using 
the indentity \eq{nerpsn} one can cast the 
equations in \eq{zerocond} entirely in terms of the 
Jacobian elliptic functions. {}From 
 \eq{nerpsn} we can obtain the following  useful identities
\bea{prelje}
\frac{\p'(u)}{\p(\omega_3) - \p(u)} =
2 \sqrt{e_1-e_3} \frac{ \cn (u \sqrt{e_1 -e_3}, \tilde k)
\dn( u \sqrt{e_1 -e_3}, \tilde k)} 
{\sn( u \sqrt{e_1 -e_3}, \tilde k)}, \cr
\frac{\p'(u)}{\p(\omega_1) - \p(u)} =
2 \sqrt{e_1-e_3} \frac{ \dn (u \sqrt{e_1 -e_3}, \tilde k)}
{\cn( u \sqrt{e_1 -e_3}, \tilde k) 
\sn( u \sqrt{e_1 -e_3}, \tilde k)}, \cr
\frac{\p'(u)}{\p(\omega_2) - \p(u)} =
2 \sqrt{e_1-e_3} \frac{ \cn (u \sqrt{e_1 -e_3}, \tilde k)}
{\dn( u \sqrt{e_1 -e_3}, \tilde k) 
\sn( u \sqrt{e_1 -e_3}, \tilde k)}. 
\eea
Substituting the above identities in \eq{zerocond} it is easy to show
using simple manipulations and indentities
involving Jacobian elliptic functions,  that the conditions reduce to the
following three equations
\bea{tranzcond}
\sum_{i} r_i \frac{\dn ( 2 u\sqrt{e_1 -e_3}, \tilde k)}{
\sn ( 2 u \sqrt{e_1 -e_3}, \tilde k)} = 0, \cr
\sum_{i} r_i \frac{\cn ( 2 u\sqrt{e_1 -e_3}, \tilde k)}{
\sn ( 2 u \sqrt{e_1 -e_3}, \tilde k)} = 0, \cr
\sum_{i} r_i \frac{1 }{
\sn ( 2 u \sqrt{e_1 -e_3}, \tilde k)} = 0.
\eea
We now use the following Landen transformations \cite{Watson} to convert the
Jacobi Elliptic functions with modulus $\tilde k$ to that 
with modulus $k$.
\bea{landen}
\sn( (1+k)u, \tilde k) &= &(1+k) \frac{\sn (u, k)}{1 + k \sn^2(u, k)}, 
\cr
\cn( (1+k)u, \tilde k) &= & \frac{\cn (u, k)\dn(u,k)}
{1 + k \sn^2(u, k)},  \cr
\dn( (1+k)u, \tilde k) &=&  \frac{1 -k \sn^2(u, k) }
{1 + k \sn^2(u, k)}.  
\eea
Note that the argument involving $u$ which occurs in 
\eq{tranzcond} is  $ 2 u \sqrt{ e_1 -e_3} = ( 1+k) u$.  
Substituting the above transformations in \eq{tranzcond} and  after  some simple manipulations we obtain the following 
equivalent 
conditions which enable us to determine $u_1, u_2, u_3$ in terms
of $u_0$ and $r_i$. 
\bea{ficonze}
\sum_i r_i \frac{1}{\sn(u_i)} = 0, \cr
\sum_i r_i \sn(u_i) =0, \cr
\sum_i r_i \frac{\cn(u_i) \dn(u_i)}{\sn(u_i)} =0.
\eea
As a simple consistency check we can show that the residues given in \eq{residue}
satisfy the equations \eq{ficonze}. Substituting the values of
$r_i$ in \eq{ficonze} and after a few manipulations, the 
three set of equations reduce to 
\bea{conscha}
\sum_i \frac{1}{\prod_{j\neq i} (z_i - z_j)}  =0, \cr
\sum_i \frac{z_i}{\prod_{j\neq i} (z_i - z_j)}  =0, \cr
\sum_i \frac{z_i^2}{\prod_{j\neq i} (z_i - z_j)}  =0, 
\eea
here $z_i = \cn(u_i)/dn(u_i)$. It can be shown by simple algebra
that the above set of equations are identities for any 
set of $z_i's$. Thus we have verified the conditions for the 
zeros \eq{ficonze}

\vspace{.5cm}
\noindent
{\emph{The Strebel differential in terms of the the $\zeta(u)$ function and 
the Strebel lengths}
\vspace{.5cm}

We now show that the Strebel differential $\sqrt{\phi(u)} du$ can also be
written in terms of the $\zeta (u)$ function. This allows one to 
perform the Strebel integral and obtain the Strebel lengths.
{}From \eq{pfnpar} we see that Strebel differential is expressed 
as a linear combination of the function
\be{baswpfc}
i r_i \frac{\wp'(u_i)}{\wp(u) - \wp(u_i)}.
\ee
This is periodic function with periods $(2\omega_1, 2\omega_2)$, with 
double zeros at the origin $u=0$ and poles at $u=\pm u_i$ with 
residues $\pm r_i$. {}From the properties of the $\zeta(u)$ function
we see that the function in \eq{baswpfc} can be written as
\be{baswpfc1}
i r_i \frac{\wp'(u_i)}{\wp(u) - \wp(u_i)} =
i r_i ( \zeta(u +u_i) - \zeta(u-u_i) - 2\zeta(u_i) ).
\ee
We now show the above indentity is true by matching the singularity
structure on both sides of the equation.
Firstly, from \eq{quasipe} we see that the right hand side of the 
above equation is also periodic with periods $(2\omega_1, 2\omega_2)$.
Since the function $\zeta(u)$ has a simple pole at $u=0$ with
residue 1, the above
combination of $\zeta$ functions has simple poles at $u= \pm u_i$ with
residues $\pm r_i$. Furthermore $\zeta(u)$ is an even function, which
implies both sides of the equation vanish at $u =0$.{}Finally,  since the 
derivative of the $\zeta(u)$ function is the $\wp(u)$ function which is 
even in $u$, the derivative of the  right hand side vanishes at $u=0$.
implying  $u=0$ is a double zero. This establishes the indentity in \eq{baswpfc1}.
Substituting \eq{baswpfc1} in \eq{pfnpar} we obtain
\be{zfnpasb}
\sqrt{\phi(u)}du = i \sum_i r_i ( \zeta( u+ u_i) - \zeta( u-u_i) - 2\zeta(u_i) ) du.
\ee
To perform the Strebel integrals in \eq{strba} and \eq{strbb} we write
the above equation in terms of the $\sigma$ functions using \eq{defsigma}.
This gives
\be{stdinsig}
\sqrt{\phi(u)} du = i \sum_i r_i \left( \frac{d}{du} \log\sigma( u+ u_i) - 
\frac{d}{du} \log\sigma( u-u_i) - 2 \zeta( u_i) \right) du.
\ee
It is now easy to perform the integrals from $0$ to $\omega_1$ and
$0$ to $\omega_2$ giving
\bea{fstblb}
a &=& \sum_i r_i [ \pi - 2 i ( \zeta(u_i)\omega_1 - \zeta(\omega_1) u_i)], \cr
b &=& \sum_i r_i [ \pi + 2 i ( \zeta(u_i)\omega_2 - \zeta(\omega_2) u_i)] .
\eea
Here we have used the quasi-periodic relations of the $\sigma(u)$ function 
given in \eq{quasisigma}

\vspace{.5cm}
\noindent
{\emph{The case of equal perimeters}}
\vspace{.5cm}

The case of equal perimeters $r_0=r_1 =r_2 =r_3 =r$ has been
studied in \cite{Belopolsky:1994bj}. Here we show that we can solve 
the conditions on the zeros \eq{ficonze} easily when the perimeters are 
equal, and the equations for the Stebel lengths \eq{fstblb} reduce to the
equations found by \cite{Belopolsky:1994bj}.
{}For the case of equal perimeters the conditions on the zeros reduce to 
\be{eqperzer}
\sum_i \frac{1}{\sn (u_i)} =0, \quad 
\sum_i \sn(u_i) =0, \quad
\sum_i \frac{\cn(u_i) \dn(u_i)}{\sn(u_i)} =0.
\ee
These conditions are  satisfied by the choice
\be{choizeeq}
u_0, \quad u_1 =u_0 + \omega_1, \quad u_2= u_0 + \omega_2,\quad u_3= u_0 +\omega_3.
\ee
Subtituting these values of $u_i$ in the conditions \eq{eqperzer}, it is easily
seen that they are satisfied using the properties in \eq{shiftellip1}. 
Writing out the Strebel differential $\sqrt{\phi(u)}$ using the above solution
for $u_i$ we obtain
\bea{strebeqres}
\sqrt{\phi(u)} du &=& i r \left( 
\frac{\wp'(u_0)}{\wp(u) - \wp(u_0)}+
\frac{\wp'(u_0+ \omega_1)}{\wp(u) - \wp(u_0+ \omega_1)} \right. \cr
&+& \left. \frac{\wp'(u_0+ \omega_2)}{\wp(u) - \wp(u_0+ \omega_2)} +
\frac{\wp'(u_0+ \omega_3)}{\wp(u) - \wp(u_0+ \omega_3)} 
\right).
\eea
It is clear from the above expression that the Strebel differential for this 
case has a periodicity with smaller periods $(\omega_1, \omega_2)$ as 
found in \cite{Belopolsky:1994bj}
Using this fact we can rewrite the differential in terms of the Weierstrass $\wp$ 
function with smaller periods, this leads to the following expression 
\be{strebdifsmpe}
\sqrt{\phi(u)} du = i r \left( \frac{\tilde\wp( u_0)}{\tilde\wp(u) -\tilde\wp(u_0)}\right),
\ee
here $\tilde\wp(u)$ is the Weierstrass function with periods $(\omega_1, \omega_2)$.
{}From the above expression one can proceed with the remaining analysis to obtain
the Strebel lengths $a, b$ as discussed here to obtain
the equations for the Strebel lengths found by \cite{Belopolsky:1994bj}.

\section{The elementary derivation of equation $(3.12)$ }

A Strebel differential with a fourth order zero can be written as
\be{strfour}
\phi(z)dz^2=C{(z-z_0)^4dz^2 \over z^2(z-1)^2(z-\e)^2},
\ee
where we have chosen to put the poles at $0,1,\infty$ and $\e$ 
using $SL(2,C)$ invariance.
Labelling the positive 
residues at these poles as $(p_1,p_2,p_3)$ and $p_0$ respectively, we see
that
\be{peqns}
{p_1\over p_3}=-{z_0^2\over \e} \quad {p_2\over p_3}=-{(z_0-1)^2\over 1-\e}.
\ee
We also have $p_0=p_1+p_2+p_3$. 
We can eliminate $z_0$ from the two equations in \eq{peqns} 
getting the quadratic equation for $\e$
\be{equad}
(p_1+p_2)\e+2\sqrt{-p_1p_3}\sqrt{\e}-(p_2+p_3).
\ee
The solution to this quadratic equation gives
\be{sqrtetaso}
\sqrt{\e}= \left( \frac{ \sqrt{p_0p_2} \pm i \sqrt{p_1p_3 }}
{p_1 + p_2} \right),
\ee
which is the same as \eq{etaso}

To connect with the expressions in
\cite{Aharony:2006th} for extremal correlators, 
we need to take $p_0 = p_t, p_1 = p_{-t}, p_2 =p_\infty, p_3 = p_1$, where
the $p$'s on the left hand side are ours and those on the right are 
theirs. Their quantity $t$ is related to $\e$ by 
\be{relett}
\eta =  \frac{(t -(-t)) (1-\infty)}{ (t-1)(-t-\infty)}
=\frac{2t}{t-1}.
\ee
According to \cite{Aharony:2006th} the 
quadratic equation which determines $t$ is 
given by
\bea{quadofer}
t^2 ( 4 + B^2 + 4A)
+ 2t ( 2B + AB) + A^2 =0 \quad {\rm where} \cr
A = \frac{p_{-t}}{p_\infty} - \frac{p_t}{p_\infty}, \quad
B = \frac{p_t}{p_\infty} + \frac{p_{-t}}{p_\infty}
\eea
Subtituting for $t$ in terms of $\eta$ in the above quadriatic equation 
we obtain the following equation for $\eta$ 
\be{eqetof}
(p_\infty + p_{-t})\eta^2 - 2\eta ( p_\infty p_t - p_1 p_{-t} ) +
(p_t -p_{-t})^2 =0
\ee
The two solutions of this equation are given in \eq{etaso} 
after making the above identifications.

\section{Details on the approximation scheme}

In this appendix we provide the details leading to the equations 
\eq{orzeqns} and \eq{eqfab} which are the basic equations to obtain
a perturbation scheme around the Y-diagram.

We start out with the conditions on the zeros given in \eq{ficonzem} and 
perform a large $z$ expansion of these equations. 
As shown in \eq{pozu} large $z$ corresponds to expansion 
aroudn $\bar u = \omega_1/2 + \omega_2$. From the 
definition of $z = \cn u/\dn u$ in \eq{ivuz} and using the identities
in \eq{sqrelelip} we can obtain the following large $z$ 
expansion of the Jacobian elliptic functions
\bea{scdexp}
\sn u &=& \frac{1}{k} \left( 1 + 
\frac{1}{2}( \frac{1}{k^2} -1 ) \frac{\epsilon^2}{z^{\prime 2} }
+ (
\frac{3}{8k^4} - \frac{1}{4k^2} - \frac{1}{8} 
) \frac{\epsilon^4}{z^{\prime 4}} +\cdots \right),
\cr
 \cn u &=& \frac{\sqrt{k^2-1}}{k}
\left( 1 + \frac{\epsilon^2}{2k^2z^{\prime 2}} + \frac{3\epsilon^4}{8 k^4 z^{\prime 4}} + \cdots \right),
\cr
\dn u &=&  \frac{\epsilon}{z'} \cn u, \cr
\frac{\cn u \dn u}{\sn u} &=&
\frac{1}{k} \left( ( k^2-1) \frac{\epsilon }{z'} + 
( \frac{k^2}{2} - \frac{1}{2k^2} ) \frac{\epsilon^3}{z^{\prime 3}} 
+ \cdots \right), 
\eea
here we have substituted $z = z'/\epsilon$ with $z'$ finite to organize the 
expansion in powers of $\epsilon$. 
Substituting these expansions
for the Jacobian elliptic functions that occur in the conditions for 
the zeros \eq{ficonzem} with
$z'_i= 1/x_i$ we obtain
\bea{zereqas}
\sum_{i=0}^3 r_i\left( 1 - \frac{1}{2} ( \frac{1}{k^2} -1)\epsilon^2 x_i^2 
+ ( \frac{-1}{8k^4} - \frac{1}{4k^2} + \frac{3}{8} ) \epsilon^4 x_i^4 + 
\cdots \right) &=& 0 , 
\cr
\sum_{i=0}^3 r_i\left( 1 + \frac{1}{2} ( \frac{1}{k^2} -1) \epsilon^2 x_i^2 
+ ( \frac{3}{8k^4} - \frac{1}{4k^2} - \frac{1}{8} ) \epsilon^4 x_i^4 + 
\cdots \right) &=& 0 , 
\cr
\sum_{i=0}^3 r_i 
\left((k^2 -1) \epsilon x_i + \frac{1}{2k^2} (k^4-1)\epsilon^3 x_i^3 \right)
&=& 0.
\eea
Adding and subtracting the first two equations and 
simplifying the third equation we obtain
\bea{simzeqnsd}
\sum_{i=0}^3 r_i  +  \sum_{i=0}^3 r_i  ( 1-k^2)^2 \frac{\epsilon^4 x_i^4}{8k^4}
&=& 0,
\cr
\sum_{i=0}^3 r_i \left( x_i^2 
+ \frac{1}{2} (1+  \frac{1}{k^2}  ) \epsilon^2x_i^4 \right) &=&0,
\cr
\sum_{i=0}^3 r_i \left(
x_i + \frac{1}{2} ( 1 + \frac{1}{k^2}) \epsilon^2 x_i^3 \right) &=&0.
\eea
Note that we have retained only the leading two terms in each of 
the equation. The above equations are the leading conditions on the 
positions of the zeros which now correspond to $x_i$. {}For later convenience
we define the following quantities
\be{defc34d}
\tilde E = - \frac{(1-k^2)^2}{8 k^4}, \qquad E = -\frac{1}{2} ( 1 +\frac{1}{k^2}).
\ee

We now find the leading expansions of the equation for the Strebel length
$a$ in \eq{fstbl} and the linear combination of lengths in \eq{fstbl} around
$\bar u$. 
Expanding
the equation for the Strebel length $a$ around
$\bar u$ we obtain
\be{stblexp}
a = A_0 \sum r_i + 
\epsilon A_1 \sum r_i x_i 
+\epsilon^2 A_2 \sum r_i x_i^2 
+\epsilon^3 A_3 \sum r_i x_i^3 
+\epsilon^4 A_4 \sum r_i x_i^4 ,
\ee
where 
\bea{defas}
A_0 &=& \pi - 2i 
(\zeta (\bar u) \omega_1 -\zeta(\omega_1) \bar u), \qquad
A_1= \frac{2i}{k} ( \zeta^{(1)} (\bar u) \omega_1 - \zeta(\omega_1)),
\cr
A_2 &=&-\frac{2i }{ 2! k^2} \zeta^{(2)}(\bar u)  \omega_1,
\qquad
A_3 = 2i ( \zeta^{(1)}(\bar u) \omega_1  - \zeta(\omega_1) ) 
\frac{1}{6k} ( 1 + \frac{1}{k^2} ) 
+\frac{2 i }{3! k^3} \zeta^{(3)} (\bar u)\omega_1, \cr
A_4 &=& - \frac{i}{3 k^2}  ( 1+ \frac{1}{k^2} ) \zeta^{(2)}(\bar u) \omega_1
- \frac{ 2i}{ 4! k^4} \zeta^{(4)} (\bar u ) \omega_1.
\eea
here the summations $\sum$ runs from $0$ to $3$, and 
\be{defdiff}
\zeta^{(n)} (\bar u) = 
\left. \frac{d^n}{du^n} \zeta(u) \right|_{ u = \bar u}.
\ee
To obtain the above expansion for $a$ we have performed a taylor series 
expansion of the equation for the Strebel length $a$ in \eq{fstbl}
about the point $\bar u$
and then substituted for $u_i$ in terms of $x_i$ using \eq{uinz}. 
Note that all the coefficients $A_{n}$ start at $O(\epsilon^0)$. 
Performing a similar expansion in \eq{comstbl} and retaining 
terms till order $\epsilon^4$ we obtain
\bea{cstblexp}
a\omega_2 + b\omega_1 &=& B_0 \sum r_i + \epsilon B_1 \sum r_i x_i 
+\epsilon^3 B_3 \sum r_i x_i^3, 
\cr
{\rm with}\; 
B_0& = &\pi \frac{\omega_1}{2}, \quad
B_1 = \frac{\pi}{k}, \quad
B_3 = \frac{\pi}{6k} ( 1 + \frac{1}{k^2}).
\eea
Here again the $B_n$'s start of at $O(\epsilon^0)$. 
We now can eliminate the combination 
$\sum r_i$, $\sum r_i x_i$ and $\sum r_i x_i^2$ using \eq{simzeqnsd} 
in \eq{stblexp} and \eq{cstblexp} to obtain the following pairs of equations
for the Strebel lengths $a$ and $b$
\bea{eqfabd}
a &=& \epsilon^3 p_1 \sum r_i x_i^3 + \epsilon^4 p_2 \sum r_i x_i^4 , \cr
a \omega_2 + b\omega_1 &=& \epsilon^3 q_1 \sum r_i x_i^3 
+ \epsilon^4 q_2 \sum r_i x_i^4.
\eea
with
\bea{defaabb}
p_1 = A_1 E + A_3, \quad p_2 = A_0 \tilde E + A_2 E + A_4, \cr
q_1 = B_1 E + B_3 , \quad q_2 = B_0 \tilde E.
\eea
{}From \eq{eqfabd} it is clear that the Strebel lengths $a, b$
begin at $O(\epsilon^3)$. This fact can also be easily verified by 
performing the scaling $z=z'/\epsilon$ in the basic equation for the
Strebel differential \eq{strebdif}.

{}From the discussion in section 5., we see that 
among the  constants in \eq{defaabb} the relevant quantity is the ratio
$p_1/q_1$. Here we  evaluate this ratio, from the definition of $p_1$ in 
terms of $A_1$ and $A_3$ we see that we 
 we first need to evaluate the derivatives
$\zeta^{(1)}(\bar u)$ and $\zeta^{(3)}(\bar u)$. 
 Using the defintion of $\zeta$ in \eq{defzetafn}, we have 
\bea{defzeeta}
\zeta^{(1)}(u) &=& -{\wp}(u), \cr
&=& \frac{ 1- 5k^2}{12} + \frac{1-k^2}{2} \frac{ \dn u}{\cn u - \dn u},
\eea
here we have used \eq{snpfn} to re-write the Weierstrass ${\wp}$ 
function in terms of the elliptic functions. 
From table 1. we see that the elliptic function $\dn(u)$ vanishes at 
  $\bar u = \omega_1/2 + \omega_2$, therefore 
we obtain
\be{valfze}
\zeta^{(1)} (\bar u) = \frac{1 - 5k^2}{12}.
\ee
Differentiating \eq{defzeeta} twice we obtain
\be{diffzeta}
\zeta^{(3)}(u) = \frac{(1-k^2)^2}{2} \frac{\dn u ( 1+ \sn^2 u) - \cn u ( 1 + k^2 \sn^2 u )}{
( \cn u - \dn u)^3 },
\ee
again since $\dn(\bar u) =0$ we get
\bea{valfzee}
\zeta^{(3)}(\bar u) &=& - \frac{( 1 -k^2)^2  ( 1 + k^2 \sn^2 \bar u)}{ 2 \cn^2 \bar u},\cr
&=& (1-k^2)k^2 ,
\eea
where we have substituted the values of the Jacobi elliptic functions at $\bar u$
from \eq{shiftelip} and table 1.
Now we can evaluate the ratio $p_1/q_1$, using the defintions
 \eq{defaabb}, \eq{defas} and \eq{defc34} 
and the equations  \eq{valfze} and \eq{valfzee} we see that
\bea{p1q1rati}
\frac{p_1}{q_1} &=& \frac{2i }{\pi} \left( \frac{1-10k^2 + k^4}{12} \omega_1 - \zeta(\omega_1) 
\right), \cr
&=& 2i \left( \frac{1-10k^2 + k^4 }{12}  \vartheta_3^2(q) 
+ \frac{1}{12} \frac{1}{\vartheta_3^2(q)}
\frac{\vartheta_1'''(q)} {\vartheta_1'(q)} 
\right).
\eea
To obtain the second line in the above equation we have used the 
relations \cite{Watson}
\be{perizetar}
\omega_1 = \pi \vartheta_3^2(q), \qquad
\zeta(\omega_1) = -\frac{\pi^2}{12\omega_1} 
\frac{\vartheta_1'''(q)} {\vartheta_1'(q)}.
\ee
In all these equations $ q= \exp( 2\pi i \omega_2/\omega_1)$, 
which implies 
\be{ratiope}
\frac{\omega_2}{\omega_1} = \frac{1}{2\pi i } \log( q).
\ee
Finally, the modulus $k$ can also be written in terms of $q$ by the 
following equation \cite{Watson}
\be{kinq}
k = \frac{\vartheta_2^2(q)}{\vartheta_3^2(q)}.
\ee

\section{Solution for the modulus $k$ in terms of the Strebel lengths.}

In this section we solve the following equation 
modulus $k$ in terms of the 
Strebel lengths $a^{(3)}$ and $b^{(3)}$.
\be{keqinab}
\frac{\omega_1 p_1}{q_1} ( \frac{\omega_2}{\omega_1} +\frac{1}{ \rho }) =1.
\ee
The above equation is basically a simple re-writing of the equation
\eq{keq}. 
{}From  now on we drop the superscript ${}^{(0)}$ which indicates the 
zeroth order terms, $\rho = a^{(3)}/b^{(3)}$. 
The strategy we use is to first write all the quantities that occur in the 
above equation in terms of the modular parameter $q$ in \eq{ratiope}, use
\eq{keqinab} to solve for $q$ in terms of the ratio of Strebel lengths $\rho$ and
then write $k$ in terms of $\rho$ using \eq{kinq}.
Since the functions involved in the equation \eq{keqinab} are transcendental
in $q$, we need to perform an expansion in $q$ to solve for $q$ in 
terms of $q$. This assumes that $|q|<1$, and for this expansion 
to be consistent it will turn out that $\rho<1$.
Substituting the expansions 
of the theta functions that occur in 
\eq{p1q1rati}, \eq{perizetar} and ,\eq{kinq} 
we obtain the following expansion 
\be{p1q1exp}
\frac{\omega_1 p_1}{q_1} = 2\pi i ( -14q + 260 q^2 - 6200 q^3 + 143368 q^4 + \cdots).
\ee
Using this expansion in \eq{keqinab} we obtain 
\bea{expqrhs}
{\rho} &=& \frac{\omega_1p_1}{q_1} \left( 1  - 
\frac{\omega_1p_1}{q_1} \frac{\log q }{2\pi i } 
   \right)^{-1},
\cr
&=& -28i\pi q + 392\pi i q^2 \log q + 520\pi i q^2 + \cdots.
\eea
We can now invert this equation and obtain $q$ in terms of $\rho$, which
is given by
\be{solinqf}
q= \frac{i}{28\pi } \rho - \frac{1}{56\pi^2} \rho^2 \log \rho -\frac{1}{2749\pi^2} ( 65-49 \log(28\pi) + i \frac{\pi}{2} ) \rho^2 + \cdots.
\ee
It is clear from this solution it is consistent to assume $|q|<1$. 
From this the leading order solution to the modulus $k$ in terms of the ratio $\rho$ is given 
by
\be{solkrho}
k =\frac{2e^{\frac{i\pi}{4}}}
{\sqrt{7\pi}} \sqrt{\rho} \left( 1 + O(\rho \log(\rho)) \right).
\ee
It is clear from the 
approach that one can obtain the solution of $k$ in terms of $\rho$ to any
accuracy that is desired.

\bibliographystyle{utphys}
\bibliography{fourpt}

\end{document}